\documentclass[sigconf]{acmart}

\usepackage{soul}
\usepackage{algorithm2e}
\usepackage{hyperref}
\usepackage{balance}
\usepackage{tabularx}
\usepackage{float}
\usepackage{multirow}
\usepackage{todonotes}
\usepackage{xcolor}
\usepackage[capitalize,noabbrev]{cleveref}
\usepackage{listings}
\usepackage{framed}
\usepackage{balance}
\definecolor{ndgold}{HTML}{D39F10}
\definecolor{oxfordblue}{rgb}{0.0, 0.13, 0.28}
\definecolor{harvardcrimson}{rgb}{0.79, 0.0, 0.09}
\definecolor{dartmouthgreen}{rgb}{0.05, 0.5, 0.06}
\definecolor{princetonorange}{rgb}{1.0, 0.56, 0.0}
\colorlet{PRINCETONORANGE}{princetonorange}
\definecolor{yaleblue}{rgb}{0.06, 0.3, 0.57}
\colorlet{YALEBLUE}{yaleblue}
\definecolor{usccardinal}{rgb}{0.6, 0.0, 0.0}
\definecolor{uclablue}{rgb}{0.33, 0.41, 0.58}
\definecolor{msugreen}{rgb}{0.09, 0.27, 0.23}
\definecolor{cornellred}{rgb}{0.7, 0.11, 0.11}
\definecolor{pomegranate}{RGB}{192, 57, 43}
\definecolor{anti-pomegranate}{RGB}{43,178,192}
\definecolor{alizarin}{RGB}{231, 76, 60}
\definecolor{anti-belize}{RGB}{185, 41, 56}
\definecolor{belize}{RGB}{41, 128, 185}
\definecolor{peter}{RGB}{52, 152, 219}
\definecolor{green}{RGB}{22, 160, 133}
\definecolor{anti-green}{RGB}{160,22,118}
\definecolor{turquoise}{RGB}{26, 188, 156}
\definecolor{pumpkin}{RGB}{211, 84, 0}
\definecolor{anti-pumpkin}{RGB}{0,22,211}
\definecolor{carrot}{RGB}{230, 126, 34}
\definecolor{wisteria}{RGB}{142, 68, 173}
\definecolor{anti-wisteria}{RGB}{99,173,68}
\definecolor{amethyst}{RGB}{155, 89, 182}
\definecolor{nephritis}{RGB}{39, 174, 96}
\definecolor{anti-nephritis}{RGB}{174,39,117}

\newcommand{\cmt}[1]{\ignorespaces}

\newcommand{\ccircle}[1]{%
\begin{tikzpicture}[enum/.style={circle,fill=ndgold,text=white,inner sep=1pt},baseline=-3pt]
  \node (node 1) at (0,0) [enum] {\scriptsize #1};%
\end{tikzpicture}}

\newcommand{\blackcircle}[1]{%
\begin{tikzpicture}[enum/.style={circle,fill=black,text=white,inner sep=1pt},baseline=-3pt]
  \node (node 1) at (0,0) [enum] {\scriptsize #1};%
\end{tikzpicture}}

\newcommand{\sys}[0]{\textsc{Crepe}}

\SetKwInput{KwInput}{Input}
\SetKwInput{KwOutput}{Output}
\SetKwInput{KwName}{Name}

\usepackage{mathtools}

\AtBeginDocument{%
  \providecommand\BibTeX{{%
    \normalfont B\kern-0.5em{\scshape i\kern-0.25em b}\kern-0.8em\TeX}}}

\copyrightyear{2026}
\acmYear{2026}
\setcopyright{cc}
\setcctype{by}
\acmConference[CHI '26]{Proceedings of the 2026 CHI Conference on Human Factors in Computing Systems}{April 13--17, 2026}{Barcelona, Spain}
\acmBooktitle{Proceedings of the 2026 CHI Conference on Human Factors in Computing Systems (CHI '26), April 13--17, 2026, Barcelona, Spain}
\acmPrice{}
\acmDOI{10.1145/3772318.3791137}
\acmISBN{979-8-4007-2278-3/2026/04}

\begin{document}


\title[\textsc{Crepe}: Mobile Screen Data Collector with Graph Query]{\textsc{Crepe}: A Mobile Screen Data Collector Using Graph Query}


\author{Yuwen Lu}
\affiliation{%
  \institution{University of Notre Dame}
  \city{Notre Dame}
  \state{IN}
  \country{USA}}
\email{ylu23@nd.edu}

\author{Meng Chen}
\affiliation{%
  \institution{University of California, Berkeley}
  \city{Berkeley}
  \state{CA}
  \country{USA}}
\email{meng.chen@berkeley.edu}

\author{Qi Zhao}
\affiliation{%
  \institution{University of Maryland, Baltimore County}
  \city{Baltimore}
  \state{MD}
  \country{USA}}
\email{qiz1@umbc.edu}

\author{Victor Cox}
\affiliation{%
  \institution{University of Notre Dame}
  \city{Notre Dame}
  \state{IN}
  \country{USA}}
\email{vcox484@gmail.com}

\author{Yang Yang}
\affiliation{%
  \institution{University of Notre Dame}
  \city{Notre Dame}
  \state{IN}
  \country{USA}}
\email{yyang1@nd.edu}

\author{Meng Jiang}
\affiliation{%
  \institution{University of Notre Dame}
  \city{Notre Dame}
  \state{IN}
  \country{USA}}
\email{mjiang2@nd.edu}

\author{Jay Brockman}
\affiliation{%
  \institution{University of Notre Dame}
  \city{Notre Dame}
  \state{IN}
  \country{USA}}
\email{jbb@nd.edu}

\author{Tamara Kay}
\affiliation{%
  \institution{University of Pittsburgh}
  \city{Pittsburgh}
  \state{PA}
  \country{USA}}
\email{}

\author{Toby Jia-Jun Li}
\affiliation{%
  \institution{University of Notre Dame}
  \city{Notre Dame}
  \state{IN}
  \country{USA}}
\email{toby.j.li@nd.edu}



\begin{teaserfigure}\includegraphics[trim=0cm 0cm 0cm 0cm, clip=true, width=\textwidth]{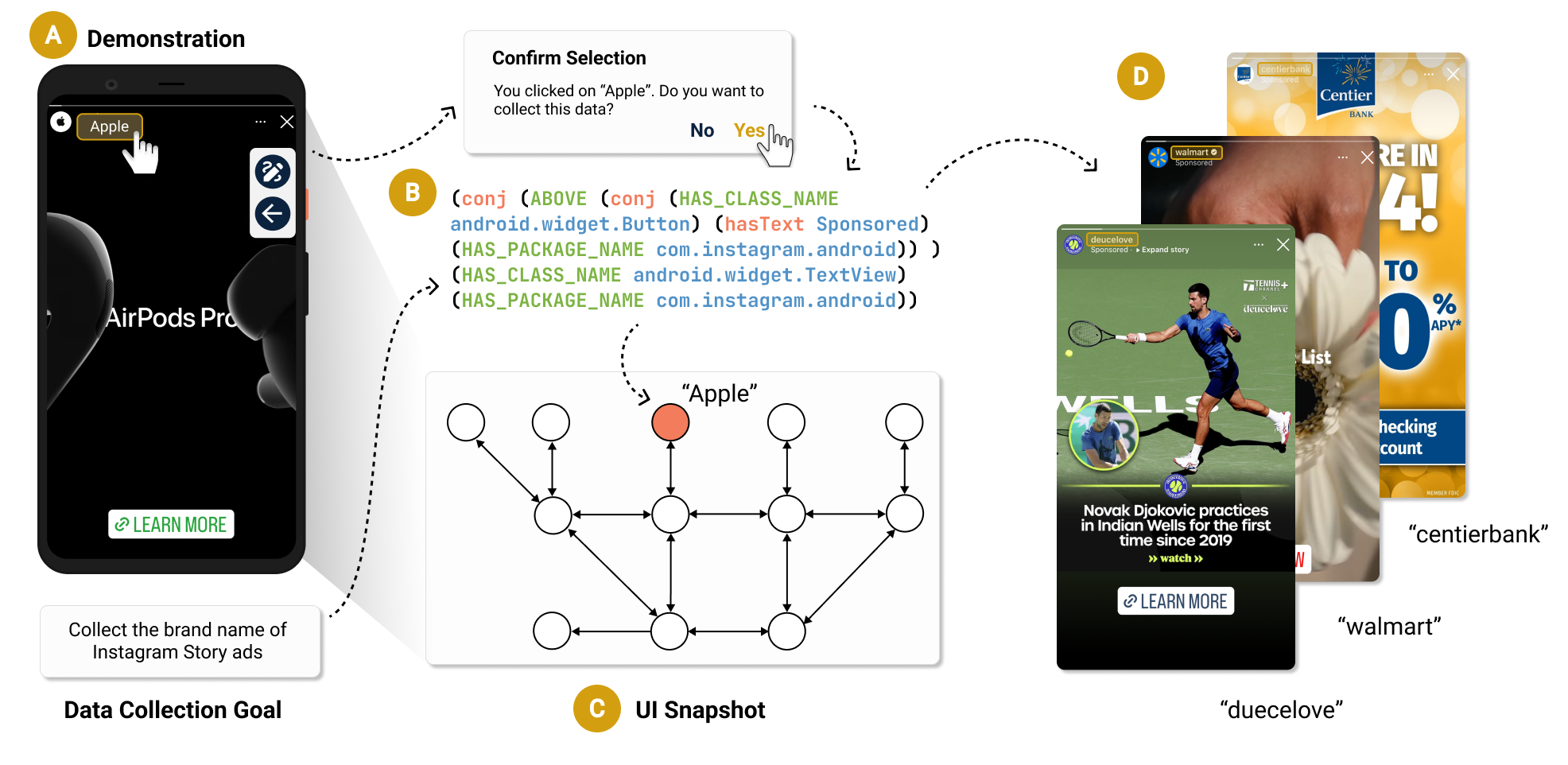}
    \caption{The \sys{} app provides a low-code solution for academic researchers to collect data displayed on mobile screens. Through a programming by demonstration paradigm, a researcher taps on the target data to collect on the screen \protect\ccircle{A}. \sys{} will automatically generate a \textit{Graph Query} we designed \protect\ccircle{B} that can accurately identify and locate the target UI element. When seeing new screens, the \textit{Graph Query} will be executed on the screen's UI Snapshot \protect\ccircle{C} and identify the UI element containing our target data. The created Graph Query will be shared with data collection study participants to collect the target data on other screens \protect\ccircle{D}.}
    \label{fig:crepe-teaser}
    \Description{The teaser figure consists of four sections: A, B, C, and D. Section A shows a screenshot from an Instagram app's Story advertisement with various UI elements like buttons, text, and images. The researcher has tapped on the "Apple" text, and a confirmation dialog appears, asking if they want to collect this data. Section B presents the automatically generated Graph Query as a structured expression, capturing the relationships between the selected ``Apple'' element and other screen entities like class names, package names, and locations. Section C displays the original UI snapshot context from the Instagram app. Section D demonstrates the execution of the generated graph query across different app UI states, successfully identifying and locating target data elements like ``centierbank'' and ``duecelove'', illustrating the precise and flexible data collection capabilities.}
\end{teaserfigure}

\begin{abstract}
Collecting mobile screen information datasets remains challenging for academic researchers. Commercial organizations often have exclusive access to mobile data, leading to a ``data monopoly'' that restricts academic research and user transparency. Existing open-source mobile data collection frameworks primarily focus on mobile sensing data rather than screen content. We present \textsc{Crepe}, a no-code Android app that enables researchers to collect information displayed on screen through simple demonstrations of target data. \textsc{Crepe} utilizes a novel Graph Query technique, which augments mobile UI structures to support flexible identification, location, and collection of specific data pieces. The tool emphasizes participants' privacy and agency by providing full transparency over collected data and allowing easy opt-out. We designed and built \textsc{Crepe} for research purposes only and in scenarios where researchers obtain explicit consent from participants. Code for \textsc{Crepe} will be open-sourced to support future academic research data collection.

\end{abstract}




\settopmatter{printfolios=true}

\maketitle

\section{Introduction}

To research on mobile information consumption and behavior, researchers often analyze screen data on users' mobile interfaces. Such data serves as a window into recommendation algorithms and how people interact with technology in everyday lives~\cite{raento_smartphones_2009}. Unlike traditional mobile sensing data that captures device states and sensor readings, screen information represents what users actually see and engage with, making it an invaluable resource for understanding the user side of interactions~\cite{brown2013iphone, rogers2017research, bentley2016thought}. 

\textcolor{black}{However, collecting mobile screen data has remained difficult for academic researchers. In practice, access to mobile app usage data is often controlled by smartphone platforms like iOS and Android, or by individual app developers who guard their user data closely~\cite{chandramouli2015insider, generosi_toolkit_2020}. Academic researchers find themselves in a challenging position: they have to either rely on limited public developer APIs, which are subject to arbitrary corporate policy changes~\cite{reddit_key_2023, barnes_twitter_2023}, or negotiate complex collaborations with commercial organizations that may compromise research independence~\cite{kramer_experimental_2014, deka_rico_2017}. This ``data monopoly'' creates barriers to empirical mobile research~\cite{jia-jun_li_bottom-up_2022}.}


\textcolor{black}{While many open-source mobile data collection frameworks exist to support academic research, most solutions focus on mobile sensing data (e.g. accelerometer, gyroscope, luminance) rather than the screen information that users see and interact with~\cite{ferreira_aware_2015, aharony_social_2011, cardone_msf_2013, teng2024tool}.} The technical challenge lies in reliably and automatically identifying which user interface screens contain target data, then locating and collecting that specific information. \textcolor{black}{Current approaches often require users to actively input data or upload screenshots manually~\cite{calacci_bargaining_2022}, or they continuously record user screens, an approach that is both inefficient and raises serious privacy concerns~\cite{krieter_can_2019, teng2024tool}.}

In this work, we ask: how can we support mobile screen data collection for academic researchers, while respecting user privacy and maintaining data quality?

\textcolor{black}{We created \sys{}\footnote{\sys{} is an acronym for ``Collector for Research Experiments of Participant Experiences''.}, a novel Android data collector that embodies a programming-by-demonstration approach to mobile screen data collection~\cite{cypher1993watch}.} At its core, \sys{} introduces Graph Query, a query language that augments mobile UI screen structures to support flexible identification, location, and collection of specific data pieces. \textcolor{black}{Graph Query helps when researchers want to collect only specific pieces of screen information, instead of full screens all the time~\cite{teng2024tool, kim2025stanford}}. With \sys{}, researchers can define what data to collect through simple, no-code demonstrations, while participants maintain full transparency and control over their data through intuitive privacy controls and real-time collection feedback.\looseness=-1

We designed \sys{} with participant agency at its center. Users see exactly what data is being collected through semi-transparent overlays and dedicated data pages, and they can leave any study at any time by simply removing the collector from their app. This approach transforms mobile data collection from an opaque, researcher-controlled process into a transparent, participant-empowered collaboration.\looseness=-1

To evaluate \sys{}, we conducted a series of three user studies with both researchers and participants, uncovering performance characteristics, limitations, and future directions in both laboratory and real-world settings.

In all, our contributions include:
\begin{enumerate}
    \item A novel query language, Graph Query, to reliably \textit{identify}, \textit{locate}, and \textit{collect} target data on UI screens; 
    \item \sys{}, the first Android data collector app to support flexible, customizable, low-code UI screen data collection using programming by demonstration (PBD)~\cite{cypher_watch_1993}\footnote{\textsc{Crepe} is open source at https://github.com/ND-SaNDwichLAB/crepe.};
    \item A series of three user studies demonstrating \sys{}'s effectiveness in empowering both researchers and participants in mobile screen data collection.
\end{enumerate}
\section{Background and Related Work}

\subsection{Understanding Human Activities via Mobile Data}
\textcolor{black}{Mobile devices, especially smartphones, are ubiquitous in people's everyday lives. Mobile applications make use of various information about user social activities, which are embedded in users' interactions, app usage patterns, and preferences. As a result, researchers in HCI and social science have been leveraging rich data gained from mobile apps to improve user experiences~\cite{deka_rico_2017, wang_enabling_2023}, uncover societal trends~\cite{vieweg_microblogging_2010, starbird_rumors_2014, starbird_pass_2010}, and understanding user behaviors and attitudes~\cite{fakhradyan_study_2021, de_predicting_2017, balaban_adolescents_2022, kramer_experimental_2014}.}

Screen data from mobile phones, i.e. information displayed on mobile screens, are particularly valuable for understanding user behavior and preferences. Screen data capture visual information and layout of the apps, providing insights into how users navigate and engage with different features. By analyzing this rich, granular data, researchers can gain a deeper understanding of user needs, preferences, and pain points, which can inform the design of more user-friendly and engaging mobile experiences~\cite{ballard2007designing}.

\subsubsection{Analyzing Behavioral Data Streams}
Researchers from diverse backgrounds, such as social science or business, mine mobile data for behavioral patterns and insights. Experience Sampling Method (ESM) is a research methodology to obtain \textit{in situ} data for constructing an understanding of subjects' daily behaviors, feelings, and thoughts of participants~\cite{van_berkel_experience_2017, yue_photographing_2014}. In the survey conducted by Berkel et al., smartphone-based ESM benefits researchers in improving data quality through validation, collecting rich multimodal data for context reconstruction, and enabling real-time data collection and analysis~\cite{van_berkel_experience_2017}. ESM can provide social scientists with combined human and sensor data in addition to background logging~\cite{raento_smartphones_2009}. Aside from sensor data, ESM questionnaires can also capture external aspects of experience, such as time, place, activities, and companion~\cite{hektner_experience_2007}. For instance, Yue et al. adopted ESM to study the function of photos. They found that photos can be used to trigger memory during follow-up interviews and as a beneficial component in data analysis~\cite{yue_photographing_2014}. In this work, our goal is to create a low-code mobile UI screen data collection tool. Our app \sys{} can facilitate academic researchers to instrument screen data collections without the need to extensively develop customized data collectors, or have public access to data APIs.

\subsection{Computational Understanding of Mobile UI Structures and Data}
Technical researchers have developed intelligent systems that achieve computational understanding of mobile user interfaces. \textsc{Sugilite}, a mobile PBD system that the execution of various tasks through user-driven multimodal direct manipulations, parses UI elements and layout information to understand the intentions contained in user activity~\cite{li_sugilite_2017}. The \textsc{Sugilite} system leverages the accessibility API of Android to encapsulate high-quality relevant behavioral data on mobile devices. UI understanding and summarization is another topic that HCI researchers are interested in. An accurate and succinct description of UI semantics can facilitate seamless interactions as well as bridge the gap between language and user interface. Prior research such as Screen2Vec~\cite{li_screen2vec_2021} and Screen2Words~\cite{wang_screen2words_2021}, adopted Machine Learning (ML) techniques to support UI design. More recent approaches have also leveraged the advance of Large Language Models (LLMs) and large multimodal models to enhance conversational interactions on mobile UI~\cite{wang_enabling_2023, yan_gpt-4v_2023}. In this work, with our proposed novel \textit{Graph Query}, we aim to create a deterministic and reliable query language that can reliably and deterministically identify and locate a piece of target information on screen. This compliments existing research on mobile UI understanding in addition to existing deep-learning-based mechanisms.



{\color{black}
\sys{} is built on Android Accessibility Service API\footnote{https://developer.android.com/guide/topics/ui/accessibility/service}, which provides a universal, framework-agnostic representation of UI elements across diverse Android app development approaches. While Android applications can be built using various frameworks including native Android Views, React Native, Flutter, Jetpack Compose, the Accessibility Service transforms all of these into a standardized accessibility tree composed of \texttt{AccessibilityNodeInfo} objects~\cite{android_accessibilitynodeinfo_2025}. This abstraction layer is required by Android guidelines to support, regardless of their internal rendering mechanisms. This enables \sys{} to collect data uniformly across native apps, hybrid apps, and web content using the same Graph Query mechanism. Applications that fail to properly implement accessibility support (e.g., canvas-based games, poorly implemented custom views) may have limited \texttt{AccessibilityNodeInfo} data, as we will discuss in Section~\ref{sec:query-limitations}.
}

\subsection{State-of-the-art Mobile Data Collection Tools} 
A variety of data collection tools and frameworks have been proposed to support academic data collection of various types of data. \textsc{Aware} is a mobile instrumental tool designed for collecting usage context through sensors on mobile devices~\cite{ferreira_aware_2015}. Rather than being a simple mobile data collector, \textsc{Aware} creates a collaborative framework for researchers to share context data with the community. \textcolor{black}{The Stanford Screenome project built a tool to take screenshots of users' phones at a set interval~\cite{robinson_screen_2017, kim2025stanford}. } \textsc{Purple Robot} is an extensible and modular development platform that supports behavioral and clinical intervention~\cite{schueller_purple_2014}. The tool enables Behavioral Intervention Technology (BIT) stakeholders to create mobile apps that collect user management, content authorship, and content delivery data, through which physician-scientists can evaluate and share data resources to increase knowledge finding and generalization. \textsc{ShiptCalculator} is a research tool designed to track and share workers' aggregated data about their pay to empower workers in offering awareness of wage transparency and advocacy to campaigns. \textcolor{black}{In the data collection process, workers send a screenshots of their pay history to \textsc{ShiptCalculator}, which parses them using OCR and stores structured data.} The system then sends a validation text to workers and allows them to explore their pay details ~\cite{calacci_bargaining_2022}. 

\textcolor{black}{However, no current tool supports the \textit{flexible} and \textit{specific} collection of data displayed on mobile UIs based on our review of previous work. A related technique is XPath, a widely-used query language for selecting nodes in XML and tree-structured documents based on structural relationships (parent/child, siblings, ancestors)~\cite{clark1999xml}. However, XPath is limited to defining document tree relationships, and is limited in expressing visual or spatial relationships as elements appear on the rendered screen. XPath also requires a fair amount of technical expertise to write the query language. In addition, a recent relevant work~\cite{teng2024tool} presents a tool that captures \textit{all} screen text from Android smartphones. While researchers can specify which apps to collect from, the tool lacks element-level specificity. Their tool collects all text displayed within those apps rather than targeting specific data pieces. This creates additional filtering burden for researchers interested in targeted within-app data. Moreover, their validation assessed data patterns and completeness without comparing against a rigorously labeled ground truth baseline. In general, either of these techniques handles data collection tasks such as \textit{``collect all Uber prices the user requested''} or \textit{``collect all Instagram ads the user saw''}\footnote{We evaluated \sys{}'s performance on these tasks in \cref{sec:eval-study-two}.}. Our design of \sys{} aims to solve these limitation, through a programming by demonstration (PBD) interaction paradigm and our design of Graph Query.}

\section{Proposed Method}

Our goal for the \sys{} data collector is three-fold. First, we seek to empower \textit{researchers} lacking programming expertise to easily create and deploy data collections. Second, we prioritize the privacy and agency of individual data collection \textit{participants}. Lastly, we also want to address the following observed \textit{technical challenges} in reliable mobile screen data collection, which hinder even technical researchers:

\begin{enumerate}
    \item The difficulty in automatically \textbf{detecting} that target data showed up on screen;

    \item The lack of precision in \textbf{locating} target data and accessing its content;

    \item The limited \textbf{generalizability} for dynamically changing data content.
\end{enumerate}

To solve these challenges, previous collectors have asked participants to either directly input target data or upload screenshots or screen recordings containing target data~\cite{calacci_bargaining_2022, krieter_can_2019}. However, these practices significantly increase participants' efforts and reduce the reliability of the collected data's quality.

In contrast, in \sys{}, we used a novel Graph Query technique to address the above challenges. Graph Query has the following characteristics:

\begin{enumerate}
    \item Triggering a data collection \textit{only} when the target data shows up; 

    \item Locating and accessing \textit{only} the target data on screen;

    \item Generalizing easily to diverse data types, especially dynamically changing data.
\end{enumerate}

For the rest of this section, we first give an overview of the experience design of \sys{}, our Android mobile data collection app. Then, we dive into the technical details of the novel \textit{Graph Query} we designed and implemented as the backbone of \sys{}.

\subsection{The \sys{} Data Collector User Experience}

\begin{figure*}
    \centering
    \includegraphics[width=\linewidth]{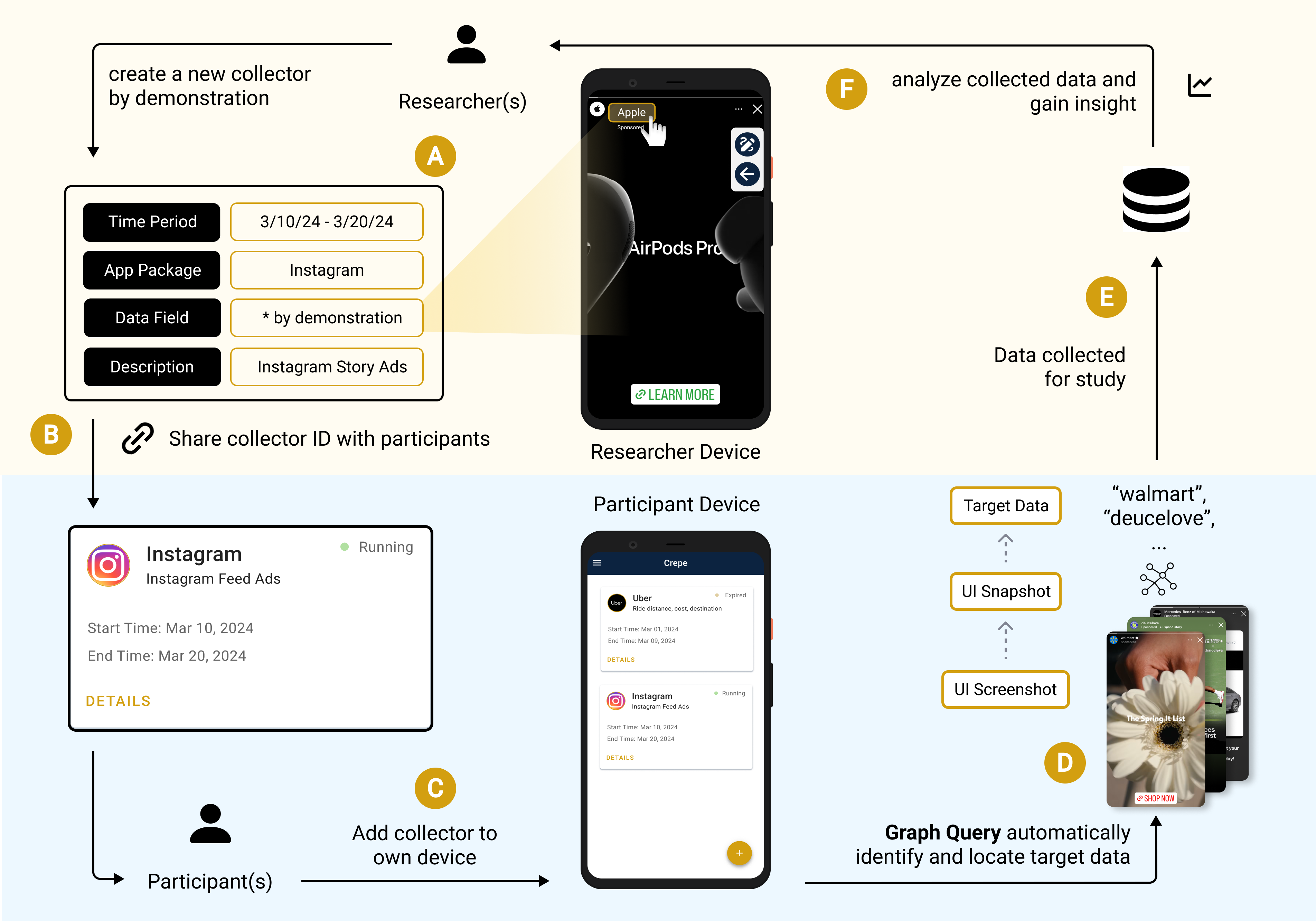}
    \caption{The workflow of \sys{} for our main user groups: data collection researchers and participants. Researchers create a new data collector by demonstration \protect\ccircle{A} and share the collector ID with participants \protect\ccircle{B}. Participants add the collector to their own devices \protect\ccircle{C}, which runs in the background to collect the specified target data \protect\ccircle{D}. Note that in step \protect\ccircle{D}, \sys{} uses Android Accessibility Service to access the view hierarchy of the current UI screen, then processes the view hierarchy, instead of directly working with screenshot images. The collected data is transmitted to a database \protect\ccircle{E} for the researcher to analyze and gain insights \protect\ccircle{F}. The colors in the background indicate each user group's experience involved in the holistic \sys{} pipeline.}
    \label{fig:workflow}
    \Description{The figure illustrates the workflow of the Crepe system with 6 labeled steps: A) The researcher creates a new data collector by specifying the time period, app package, data field and description through a graphical interface on their device. B) The researcher shares the collector ID with participants. C) Participants add the collector to their own devices by entering the shared collector ID into the Instagram app running on their phone. The collector details like start/end time are shown. D) On the participant's device, the Crepe app automatically identifies and locates the target UI data elements (like ``walmart'', ``deucelove'') using Graph Query, based on the UI screenshot. E) The target data elements collected by Crepe on participant devices get transmitted to a database. F) The researcher analyzes the collected data from the database to gain insights for their study. Arrows show the flow of information between the researcher device, participant devices, and the data collection database.}
\end{figure*}

The \sys{} data collector has two main target user groups: \textit{researchers} who want to instrument a mobile data collection study, and \textit{study participants} who want to contribute mobile data to ongoing data collections\footnote{We intend \sys{} to be used for research purposes only where participants give their explicit consent to researchers before data collection begins.}.

Researchers have two main tasks when using \sys{} (\cref{fig:workflow}): creating a collector \protect\ccircle{A}, and sharing collectors with study participants \protect\ccircle{B}. A researcher can use \sys{}'s graphical user interfaces to define the start and end dates of a collector, as well as the target app to collect data from. 

The most important yet trickiest step in defining a data collector is \textit{specifying the target data} on screen to collect. In \sys{}, our design goal is to minimize the complexity of this specification process for researchers, by only asking them to \textit{tap on} the screen data to collect in the target app, and \textit{describe} their intention to collect these data (\cref{fig:workflow} \protect\ccircle{A}). \sys{} automatically translates these two pieces of information into a formal, executable \textit{Graph Query}, which includes information about the target data's parent app package, component type, and relations with other UI entities on screen. This \textit{Graph Query} is the core of a data collector and will be used to reliably identify and precisely locate target data on the \textit{study participants'} devices. This specification process follows a programming-by-demonstration paradigm~\cite{cypher_watch_1993} and provides a simple, no-code experience for researchers to specify the target data. (The technical detail will be further explained in the following subsection on \textit{Graph Query}.) After the target data are specified and descriptions are added, a collector is created. Afterward, the researcher can copy and share the unique ID of this collector with participants \protect\ccircle{B}. The collector will run on the participants' device for the specified data collection periods. Afterwards, the researcher can receive the data collected in a self-defined database (E) and analyze it to uncover research insights (F). Our initial implementation of \sys{} transmits collected data to a pre-configured Google Firebase Realtime Database that encrypts data both in transit and at rest\footnote{https://firebase.google.com/support/privacy}. We plan to support researchers' customized configuration of their own database servers in the future. 

For study participants, they can easily add a collector to \sys{} by entering the unique collector ID shared by the study's researchers \protect\ccircle{C}, using \sys{}'s graphical user interfaces. After granting \sys{} with the necessary Android Accessibility Service permission, the added collector(s) will start automatically running in the background of the participants' devices (D). Guided by the \textit{Graph Queries} associated with the participants' added collectors, \sys{} will selectively seek the appearance of target data \textit{exclusively} when the designated target apps are active. Once the target data is detected, \sys{} will scrape and transmit the relevant information to the researchers (E). The participant will be able to monitor the active status of the collector(s) they have joined in \sys{}. To drop out of a study, participants can simply delete a collector from their \sys{} app, and the collector will no longer run in the background.

\begin{figure*}
    \centering
    \includegraphics[width=\linewidth]{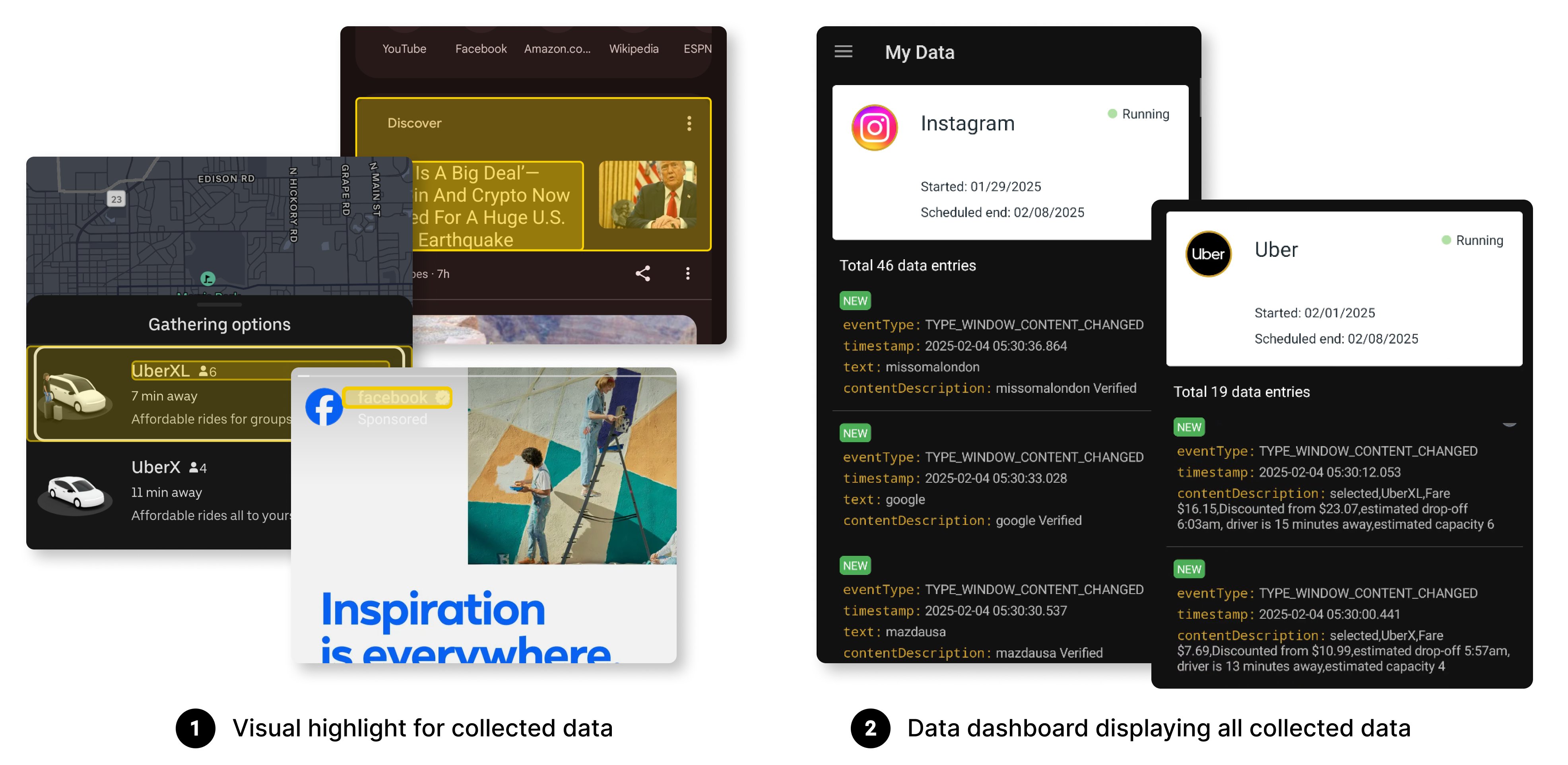}
    \caption{Two design features that enhance data collection transparency for system users (data contributors). When the system operates in the background, a yellow highlight appears over the collected content at each time of collection (\protect\blackcircle{1}). Additionally, we designed and implemented a page that displays all collected data organized by collector, allowing users to review their contribution history and details (\protect\blackcircle{2}).}
    \label{fig:collection_transparency}
    \Description{This image displays two key transparency features of a data collection system interface: The left side shows multiple overlapping interface elements, including a map in the background, an Uber ride selection menu, a news article with a headline about cryptocurrency, and a Facebook promotional card. A yellow highlight surrounds the active content being collected, drawing attention to what data is being captured. This section is labeled ``1 Visual highlight for collected data.'' The right side shows a data dashboard interface titled "My Data" that provides users with a detailed log of their collected data. The dashboard displays information for two data collectors: Instagram and Uber. For each collector, it shows when collection started (January 29, 2025 for Instagram, February 1, 2025 for Uber), the scheduled end date (February 8, 2025 for both), and the total number of data entries (46 for Instagram, 19 for Uber). The dashboard also displays individual data collection events with timestamps, event types, and content descriptions, including specific details about Uber rides such as fare information, pickup times, and driver proximity. This section is labeled ``2 Data dashboard displaying all collected data.'' Together, these features provide transparency by highlighting data as it's collected and giving users access to a comprehensive record of their contributed data.}
\end{figure*}

\paragraph{Data collection transparency}
Ensuring a transparent communication of data collection is critical in our design of \sys{}. We ensure participants have the \textit{full control} over their participation status and data contributions using two main measures. First, we adopt an ``opt-in'' participation mechanism, where participants actively add collectors to their \sys{} app to join data collections. We intend this to be combined with the explicit consents that researchers must obtain before each data collection study starts. Second, each participant can opt out of any collection, at any time, by deleting the collector from their \sys{} app, which will immediately terminate the collection on their mobile devices. 

We also prioritized data collection transparency in \sys{} design \cref{fig:collection_transparency}. Every time data is collected from the user's screen, a semi-transparent yellow highlight will appear over the collected data every time (\protect\blackcircle{1}). We also designed a dedicated screen to show the users all of the data collected (\protect\blackcircle{2}). 
We will also continuously release new privacy-centric features, such as daily or weekly push notifications summarizing data contributions, and the ability to request all their data be deleted when they withdraw from a study.

\subsection{Graph Query}
\subsubsection{Building blocks of Graph Queries: screen entities and their relations}
\label{sec:graph-query-building-blocks}


\begin{table*}[t]
\centering
\small 
{\color{black}
\begin{tabular}{p{4cm} l l p{5cm}}
\toprule
\textbf{UI Element Attribute} & \textbf{Category} & \textbf{Deprioritized} & \textbf{Description} \\
\midrule
\multicolumn{4}{l}{\textit{Accessibility API metadata}} \\
\texttt{hasPackageName} & Implementation & No & Ensure query runs in target app only \\
\texttt{hasClassName} & Implementation & No & Filter by widget type (Button, TextView) \\
\texttt{hasText} & Semantic & No & Locate element by displayed text \\
\texttt{hasContentDescription} & Semantic & Slightly & Use accessibility labels for identification \\
\texttt{isClickable / isEditable / isScrollable} & Implementation & Slightly & Filter by interaction capabilities \\
\texttt{hasViewId} & Implementation & Yes & Identify by Android resource ID (may change, thus deprioritized) \\
\texttt{hasScreenLocation} & Implementation & Yes & Fixed coordinates (fragile across devices) \\
\midrule
\multicolumn{4}{l}{\textit{Semantic Content Relations (Computed)}} \\
\texttt{containsMoney} & Semantic & No & Extract prices, wages (e.g., turning ``\$12.50'' string into a numerical value of 12.50) \\
\texttt{containsDate / containsTime} & Semantic & No & Parse temporal information for tracking \\
\texttt{containsPhoneNumber / containsEmailAddress} & Semantic & No & Extract contact information \\
\texttt{containsNumber} & Semantic & Slightly & Generic numeric data extraction \\
\texttt{containsPercentage / containsTemperature} & Semantic & Slightly & Domain-specific numeric parsing \\
\midrule
\multicolumn{4}{l}{\textit{Hierarchical Relations (Computed)}} \\
\texttt{hasParent / hasChild} & Hierarchical & No & Navigate view hierarchy (e.g., list items) \\
\texttt{hasParentText / hasChildText} & Hierarchical & No & Identify element by parent/child content \\
\texttt{hasSiblingText} & Hierarchical & Slightly & Use adjacent elements as context \\
\texttt{hasListOrder} & Hierarchical & Slightly & Locate nth item in scrollable lists \\
\midrule
\multicolumn{4}{l}{\textit{Spatial Relations (Computed)}} \\
\texttt{above / below / left / right} & Spatial & No & Locate relative to landmark elements \\
\texttt{near / nextTo} & Spatial & Slightly & Proximity-based identification \\
\bottomrule
\end{tabular}
}
\caption{\textcolor{black}{\textbf{A list of the main screen element metadata and relations we used to construct Graph Queries.} Accessibility API metadata are directly retrieved through Android Accessibility API, while the others were computed by \sys{}. These computed attributes enable our Graph Queries to express the semantic, structural, and spatial information of target screen data. We deprioritize relations that are less generalizable across screens, such as fixed screen coordinates and Android Resource IDs, but still use them as necessary fallbacks. When constructing candidate Graph Queries, we first discard queries that cannot uniquely identify only the target data on the screen, then select the top Graph Query based on the attribute priorities it contains (see \cref{fig:query_details}, \protect\ccircle{2}).}}
\label{table:screen-metadata}
\Description{A table listing main screen metadata and element relations available in the Graph Query system, organized into four categories: Accessibility API metadata, Semantic Content Relations, Hierarchical Relations, and Spatial Relations. Each relation is annotated with its category type, priority ranking (High/Medium/Low), and example use cases.}
\end{table*}

\begin{figure*}
    \centering
    \includegraphics[width=\linewidth]{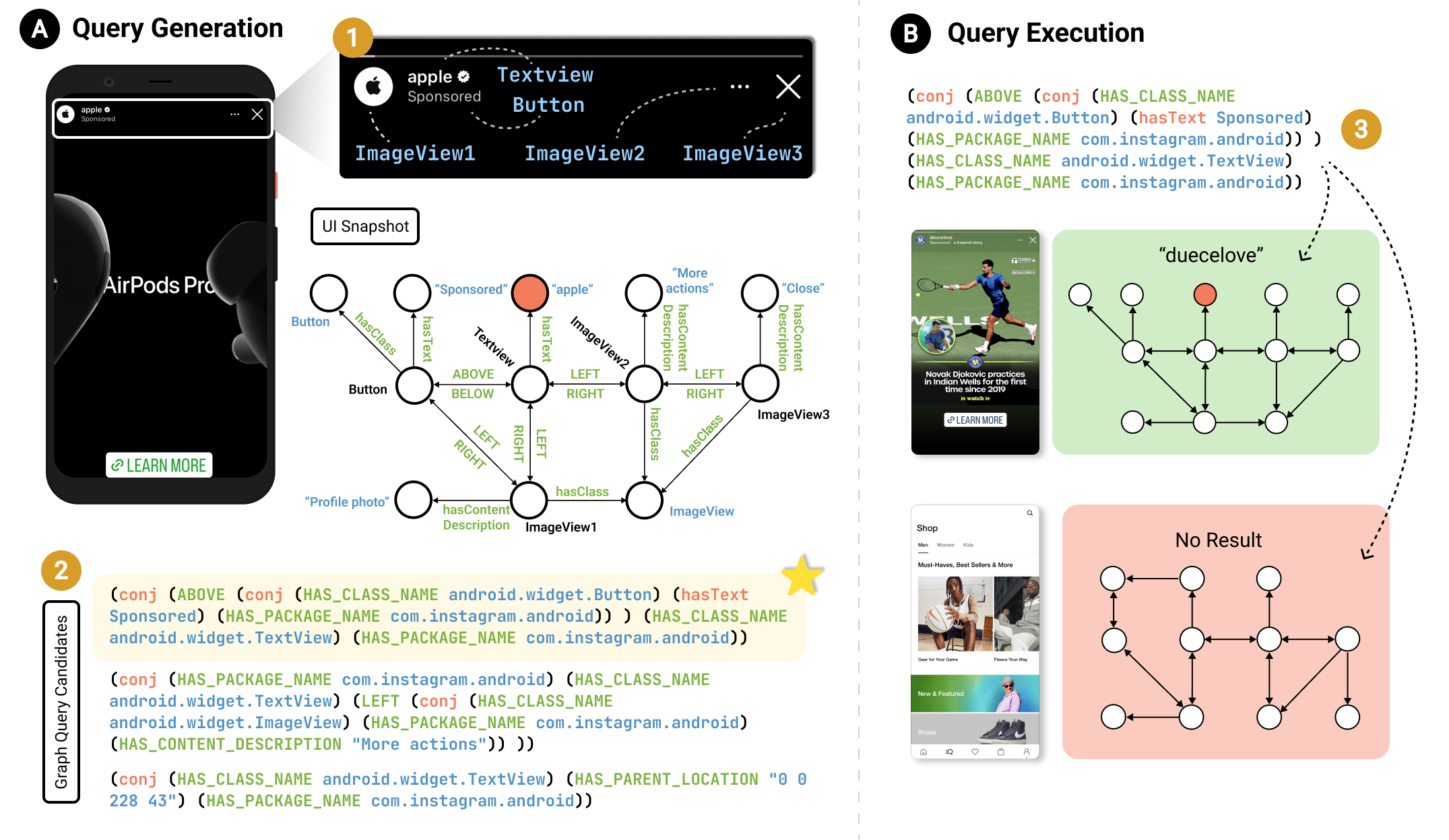}
    \caption{The detailed process of Graph Query generation and execution in \sys{}. Section \protect\blackcircle{A} shows how to generate a Graph Query: \sys{} first uses Android Accessibility Service to access the view hierarchy of a UI screen \protect\ccircle{1} and augment it into a UI Snapshot graph, using UI elements' characteristics and their relations. To identify the target UI element within the UI Snapshot graph, \sys{} combines a set of UI element characteristics on this UI Snapshot to uniquely identify the target UI element on this UI Snapshot. We put these characteristics together to construct our Graph Query \protect\ccircle{2} and rank them based on their flexibility. In Section \protect\blackcircle{B}, step \protect\ccircle{3} depicts the execution of a chosen Graph Query on two new screens, one containing the target data and one not. The set of unique characteristics of the target UI element ensures we only successfully collect the target data in ``duecelove'', but not on the other screen.}
    \label{fig:query_details}
    \Description{The figure consists of two main sections, A and B, presented side by side. Section A, labeled ``Query Generation'', displays a UI snapshot from a mobile app with various GUI elements like buttons, images, and text. It shows the process of creating Graph Queries by combining the relations between these screen entities. Section B, titled ``Query Execution'', demonstrates the execution of a selected Graph Query on different UI snapshots. The query successfully identifies and locates the target element ``duecelove'' on one snapshot but fails to find a match on another screen. The figure uses visual cues like arrows and graphs to illustrate the flow of query generation and execution.}
\end{figure*}

We develop a novel Graph Query in \sys{} for accurate and consistent extraction of various types of screen information (\cref{fig:query_details}). \textcolor{black}{The intuition behind Graph Query is to \textit{uniquely} identify the element(s) that contain the target data using (1) the UI elements' own attributes and (2) relations with other elements on screen. Graph Query builds on top of the XML mobile screen hierarchy from Android Accessibility Service, similar to the DOM tree of web HTML. This mobile screen hierarchy contains \textit{Views} on the current screen (\cref{fig:query_details}, \protect\ccircle{1}), each of which contains rich properties including content, screen location, etc. }

\textcolor{black}{First, we enhance the default mobile screen hierarchy provided by Android Accessibility Service into an augmented \textit{UI Snapshot}, which contains additional relations between screen entities (\cref{fig:query_details}, UI Snapshot section). We defined a screen entity to be either an UI element (i.e., Android View) or an attribute of the UI element, in data representations such as strings or integers. The relations between these screen entities include: (1) UI elements to their own Android implementation attributes (e.g., \texttt{hasClassName}, \texttt{hasScreenLocation}, \texttt{isEditable}); (2) UI elements to their semantic content (e.g., \texttt{hasText}, \texttt{containsEmailAddress}, \texttt{containsMoney}, \texttt{containsDate}); (3) hierarchical relations between UI elements in the XML screen hierarchy (e.g., \texttt{hasParent}, \texttt{hasChild}, \texttt{hasSibling}); and (4) spatial relations between UI elements based on their rendered screen locations (e.g., \texttt{right}, \texttt{left}, \texttt{above}, \texttt{below}, \texttt{near}).}

\textcolor{black}{Table~\ref{table:screen-metadata} provides a comprehensive list of the main screen metadata and element relations used to construct Graph Queries, organized by category, deprioritization status, and example use cases. Spatial relations are computed from UI elements' bounding box coordinates, enabling queries like ``the price text left of the quantity input field''. Semantic relations automatically parse structured data from text (e.g., extracting \$12.50 as a numeric price). In the future, this set of relations can be \textit{extended}, i.e. researchers can define custom relations for domain-specific needs (e.g., \texttt{hasStarRating} for review collection, \texttt{containsHashtag} for social media analysis). The priority of attributes are empirically decided through trial and error, with the goal as getting the most generalizable Graph Query candidate that can flexibly and accurately collect the target data across different screens (more details in \cref{sec:user-intent-disambiguation}.}

These relations comprehensively encapsulate the various relations and semantic contents tha are useful for identifying and locating a date on screen. With such additional information, an \textit{UI Snapshot} is a collection of subject-predicate-object triples denoted as (\textit{s}, \textit{o}, \textit{p}), where \textit{s} and \textit{o} are two entities and the \textit{p} is the relation between \textit{s} and \textit{o}. \looseness=-1


\begin{table}[h!]
\centering
\begin{tabular}{c c}
\hline
Expression & Rule \\
\hline
E & $\rightarrow$ e; \\
E & $\rightarrow$ S; \\
S & $\rightarrow$ (join r E); \\
S & $\rightarrow$ (and S S); \\
T & $\rightarrow$ (ARG\_MAX r S); \\
T & $\rightarrow$ (ARG\_MIN r S); \\
Q & $\rightarrow$ S | T; \\
\hline
\end{tabular}
\caption{\textbf{Context-free grammars (CFGs) for constructing Graph Query in \sys{}.}  \textit{Q} denotes the initial non-terminal symbol, terminal \textit{e} denotes a GUI object entity, and terminal \textit{r} denotes a relation. Other non-terminal symbols are employed for intermediary steps in the derivation process. We used this set of CFGs to automatically construct formal Graph Queries from the screen entity relations in UI Snapshots.}
\label{table:grammar_rules}
\Description{This figure presents a table outlining the context-free grammar (CFG) rules used to construct Graph Queries in the Crepe system. The table has two columns: the left column lists grammar expressions, and the right column shows their corresponding production rules.

E can be replaced by either e; or S; S can be replaced by (join r E) or (and S S). T can be replaced by (ARG\_MAX r S) or (ARG\_MIN r S). Q, the initial non-terminal symbol, can be replaced by either S or T
}
\end{table}

\subsubsection{Automatic generation of Graph Queries}
\label{sec:query-creation}
\textcolor{black}{When a researcher demonstrate the target data to collect by tapping on it (e.g. tapping on ``apple'' on the screen in \cref{fig:query_details} Step \protect\ccircle{1}), we locate the associated UI element in the \textit{UI Snapshot}, and utilize a unique set of its characteristics in the \textit{UI Snapshot} to generate Graph Query candidates (\cref{fig:query_details}, Step \protect\ccircle{2}). 
}
The Graph Query can be connected via 3 logical operators: \texttt{conj} (and), \texttt{or} (or), and \texttt{prev} (previous). We defined a set of context-free grammars, summarized in Table ~\ref{table:grammar_rules}, to construct formal Graph Queries based on screen entity relations in UI Snapshots.

\textcolor{black}{The context-free grammar (CFG) in Table~\ref{table:grammar_rules} defines how Graph Queries are systematically constructed from screen entity relations in UI Snapshots, similar to how CFGs are used in compilers to parse programming language syntax~\cite{alfred2007compilers}. The base case \textbf{E → e} represents a single UI entity, while the rule \textbf{S → (join r E)} combines an entity with a relation (e.g., \texttt{(hasText ``Sponsored'')}), and \textbf{S → (and S S)} chains multiple conditions using logical operators (\texttt{conj}, \texttt{or}) or spatial/hierarchical operators (\texttt{above}, \texttt{hasParent}). The aggregation rules \textbf{T → (ARG\_MAX r S)} and \textbf{T → (ARG\_MIN r S)} enable selecting entities based on numeric properties, such as \texttt{argmin hasListOrder} to find the first list item. These grammar rules ensure any valid Graph Query, from simple matches like \texttt{(isClickable true)} to complex nested expressions like \texttt{(above (conj (hasText ``Sponsored'') (hasClassName ``Button'')))}, can be systematically constructed, parsed, and executed across different screens.} As a result, \textit{Graph Queries} encapsulate combinations of the screen entity relations for the target UI element and \sys{} creates combinations can uniquely identify \textit{only} the target UI element on screen (\cref{fig:query_details}, Step \protect\ccircle{2}). The combinations of screen entities are empirically set through experimentation. All Graph Queries will have \texttt{hasPackageName} and \texttt{hasClassName} to establish the context. In most cases, more than one Graph Query can uniquely locate the target UI element on the screen.

\subsubsection{Graph Query Generation Walkthrough}

To illustrate the above technical steps in detail, here we map out the steps involved in generating a Graph Query through a detailed example, in complement to \cref{fig:query_details}.

\vspace{1em}

\begin{figure*}[t]
\raggedright

\begin{framed}

\noindent\textbf{Example: Collecting Instagram Ad Advertiser Names}
\smallskip

\noindent\textbf{Scenario:} A researcher wants to collect advertiser names from Instagram story ads (e.g., ``apple'', ``nike''). The advertiser name appears as text above a ``Sponsored'' button.
\smallskip

\noindent\textbf{Step 1 - Accessibility Service captures screen:} When the researcher taps on ``apple'', Android Accessibility Service provides the view hierarchy, containing relevant nodes including:
\begin{itemize}
    \item TextView (text=``apple'', className=``android.widget.TextView'', bounds=[10,100][200,150])
    \item Button (text=``Sponsored'', className=``android.widget.Button'', bounds=[10,160][200,200])
\end{itemize}
\smallskip

\noindent\textbf{Step 2 - Augmentation into UI Snapshot:} \sys{} converts this into triples, adding computed spatial relations:
\begin{itemize}
    \item \texttt{(node\_1, hasText, ``apple'')}
    \item \texttt{(node\_1, hasClassName, ``android.widget.TextView'')}
    \item \texttt{(node\_1, hasScreenLocation, Rect[10,100][200,150])}
    \item \texttt{(node\_2, hasText, ``Sponsored'')}
    \item \texttt{(node\_2, hasClassName, ``android.widget.Button'')}
    \item \texttt{(node\_1, above, node\_2)} \textit{← computed from UI element bounding boxes}
\end{itemize}
\smallskip

\noindent\textbf{Step 3 - Query Generation:} \sys{} creates candidate queries that uniquely match \texttt{node\_1}, using the Context Free Grammar defined in \cref{table:grammar_rules}:
\begin{itemize}
    \item Query 1: \texttt{(above (conj (hasText ``Sponsored'') (hasClassName ``Button'')))}
    \item[] \hspace{1em} in natural language means: ``\textit{Text above the button that says `Sponsored'}''
    \item Query 2: \texttt{(conj (hasScreenLocation Rect[10,100][200,150]) (hasText ``apple''))}
    \item[] \hspace{1em} in natural language means: ``\textit{Text `apple' at screen location (10,100)}''
\end{itemize}
\smallskip
\noindent Query 1 is ranked higher due to better generalizability (spatial relations vs. fixed coordinates). When executed on new screens, Query 1 successfully collects ``nike'', ``adidas'', etc., while Query 2 would fail if advertisers appear at different screen locations. This example also illustrates an important idea in Graph Query: for dynamically changing content, it is often easier to find an ``anchor point'' that does not change often (in this case, the button with text ``Sponsored '').
\end{framed}

\caption{A walkthrough example demonstrating how \sys{} constructs a UI Snapshot and generates Graph Queries. Starting from raw accessibility data, the system augments the UI structure with spatial relations and produces candidate queries that can locate target data across different screens. Queries that are more generalizable, for example, those that are based on more stable anchor points (e.g., the ``Sponsored'' button) are ranked higher.}
\Description{A framed text box presenting a step-by-step example of collecting Instagram ad advertiser names using the Crepe system. The example begins with a scenario where a researcher wants to collect advertiser names appearing above a Sponsored button. Step 1 shows how Android Accessibility Service captures the view hierarchy, returning nodes for a TextView containing "apple" and a Button containing "Sponsored" with their bounding box coordinates. Step 2 demonstrates the augmentation into a UI Snapshot, converting the hierarchy into semantic triples such as (node\_1, hasText, "apple") and (node\_1, above, node\_2), where spatial relations are computed from bounding boxes. Step 3 illustrates query generation, producing two candidate queries: Query 1 uses spatial relations to find text above the Sponsored button, while Query 2 uses fixed screen coordinates. The example concludes by explaining that Query 1 ranks higher due to better generalizability across different screens, and emphasizes the importance of finding stable anchor points for dynamically changing content.}  
\label{fig:insta_example}

\end{figure*}

\vspace{1em}

\textcolor{black}{Note that the above query examples are simplified for easier understanding. Please refer to the Graph Queries in \cref{table:grammar_rules} for real example Graph Queries.}

\subsubsection{User Intent Disambiguation: Selecting the Best Graph Query}
\label{sec:user-intent-disambiguation}
As described above, for each target data, \sys{} generates a few Graph Query candidates that can each uniquely match the target data. However, some query candidates are more generalizable to other screens: for example, to collect Instagram story ads, a query that collects the text above ``Sponsored'' is more likely to collect all target data, than a query that collects text that shows up at a specific screen location (the first and third queries in \cref{fig:query_details}, Step \protect\ccircle{2}). \textcolor{black}{As a result, we empirically defined UI element attribute priorities and used a set of heuristics to rank the generalizability of generated queries (\cref{table:screen-metadata}). Yet, from our user studies, we realized that it is best to adopt a human-in-the-loop process, engaging the researcher in this selection process.}

Through iterative design with researchers (details in \cref{sec:eval-one}) and taking inspirations from past research~\cite{li2018appinite}, our final design presents the top Graph Query candidates in natural language phrases for the user to select from. We translate Graph Queries into natural language phrases, such as ``the UI element above the Button that says \textit{Sponsored}'', to show to the user instead of its original query language form (\cref{fig:query_details}, Section \protect\ccircle{2}, option 1). We translated the Graph Queries so that the researchers, especially the non-technical ones, are not exposed to unnecessary technical details of \sys{}. We translate the Graph Queries using a large language model\footnote{We used GPT-3.5-turbo in our implementation due to its ease of access, short inference time, and empirically high-quality outputs observed on our task. Our prompt can be found in Appendix \ref{appendix-prompt}.}. This ensures the selected Graph Query best reflects the researcher's data collection goals.




\subsubsection{Graph Query execution on UI Snapshots: retrieving the target data}
After a Graph Query is created, it can be used to collect target data on any \textit{UI Snapshot} with elements sharing the same set of entity relations. \textcolor{black}{In the background, \sys{} runs the Graph Query every time a UI Snapshot is updated from changes in screen content (\cref{fig:query_details}, Section \protect\blackcircle{B}). We limit the Graph Query to only run in the target app package, in order to effectively reduce unnecessary battery usage of \sys{}.} After identifying and locating the target UI element on screen (Step \protect\ccircle{3}), \sys{} can collect the target data to collect through the element's attributes such as text and content description. 

\subsubsection{Summary}
Graph Query showcases three major strengths in screen data collection. First, the data collection mechanism using Graph Query is \textit{fully deterministic} and thus much more \textit{reliable} than alternative solutions like using Optical Character Recognition (OCR) or deep-learning based detection models. Second, Graph Query provides much more flexibility in identifying and locating target UI elements through their implementation attributes, semantic content, hierarchical screen structure position, and screen location relations (\cref{sec:graph-query-building-blocks}). For instance, a researcher can utilize the pre-defined \texttt{hasPrice} relation to collect Uber drivers' task payment information, a common task on gig worker support platforms~\cite{calacci_bargaining_2022}. In the future, developers can also extend existing entity relation types for more customized data collection tasks. Lastly, by utilizing Android's Accessibility Service, using Graph Query can easily integrate the collection of user interaction data, a feature included in \sys{} yet unsupported in most vision-based data collection tools. \textcolor{black}{In the future, \sys{} can be extended to collect more rich media information such as screenshots under the study participants' consent~\cite{android_accessibility_screenshot}.}

{\color{black}
\subsubsection{Limitations and Extensibility}
\label{sec:query-limitations}
Graph Query's relational encoding provides robustness to many challenges in mobile data collection, as long as the semantic, hierarchical, and spatial structure defined in the query is preserved. Our main assumption is that even across app updates and redesigns, major semantic information and screen structures are likely to be preserved for consistent user experience. 
However, we acknowledge that certain app redesign and localization changes might limit our Graph Query's generalizability. Table~\ref{table:limitations} summarizes scenarios where Graph Query faces challenges and potential mitigation strategies. In general, the Graph Query language is designed as a set of \textit{composable primitives} rather than fixed templates for data collection. In fact, in our evaluation study (\cref{sec:edge-cases}), we encountered the limitation of changing anchors for Instagram stories. We were able to work around it by demonstrating the target data in multiple potential layouts.

\begin{table}[h!]
\centering
\small
{\color{black}
\begin{tabular}{p{3cm} p{\dimexpr\linewidth-3cm-4\tabcolsep-3\arrayrulewidth}}
\toprule
\textbf{Limitation} & \textbf{Description and Potential Workarounds} \\
\midrule
Localization that changes layout direction & Graph Queries using spatial attributes such as \texttt{left}/\texttt{right} may break; researchers need to add multiple Graph Queries to cover different layout scenarios if they consider cross-cultral data collection needs \\
\midrule
Broad, underspecified collection needs & For example, ``collecting all ads users see on their phone''. Researchers must demonstrate each data to collect specifically; \sys{} then creates separate queries for each demonstrated scenario \\
\midrule
Dynamically generated content with no stable anchors & For example, infinite scroll feeds with no headers. Researchers may use \texttt{hasListOrder} with \texttt{argmin}/\texttt{argmax} to systematically collect items; may require demonstration of multiple examples to establish pattern \\
\midrule
Canvas-based UI elements (e.g., game graphics, custom-drawn charts) & Android Accessibility Service can be limited for canvas content; researcher can extend \sys{} to include screenshot capabilities to handle these scenarios~\cite{android_accessibility_screenshot} \\
\midrule
Non-text rich media analysis (images, videos, audio) & Screenshot API available~\cite{android_accessibility_screenshot} for image capture; can extend with computer vision or audio transcription modules for content analysis \\
\bottomrule
\end{tabular}
}
\caption{\textcolor{black}{\textbf{Limitations of Graph Query and potential workarounds.} This table shows a list of potential cases where Graph Queries might not handle well. For cases where one Graph Query cannot comprehensively capture all potential structures, we designed \sys{} to support multiple Graph Queries, allowing researchers to handle localization changes (e.g., RTL/LTR layouts) and varying app designs across A/B testing or different devices.}}
\label{table:limitations}
\Description{A table outlining current limitations of the Graph Query system and practical workarounds. Limitations include localization issues, broad collection needs, canvas-based UIs, dynamic content without anchors, and rich media analysis.}
\end{table}
}

\subsection{Implementation}
\sys{} was implemented using Android, Google Firebase Realtime Database, OAuth, and OpenAI API. Note that we only used OpenAI large language model for Graph Query translation (\cref{sec:user-intent-disambiguation}) and the whole data collection process does not involve any deep learning. \textcolor{black}{The app was developed in Java using Android Studio and is compatible with mobile devices running Android 9.0 or above (around 95.4\% of the Android mobile phone market as of February 2025~\cite{android_distribution_chart}). Firebase Realtime Database was selected due to its high-standard data encryption both in transit and at rest, real-time data synchronization, and reliable user authentication. However, we recognize that Firebase might not serve all data collection and compliance needs; since we will open source \sys{}, researchers have the full customizability of desired database provider and storage location (see deployment options below).} 

{\color{black}To implement the query language, \sys{} uses the Android Accessibility Service and captures Accessibility Events as a result of changes in the content of the screen. The captured Accessibility Event contains the new UI screen hierarchy\footnote{https://developer.android.com/reference/android/view/accessibility/AccessibilityEvent}. Android Accessibility Service gets triggered frequently as screen content changes (potentially many times per second), so it is necessary to avoid duplicate query execution and data collection, for optimized battery consumption and data storage. In our implementation, we prevent collecting duplicate data through two main mechanisms: (1) content-based deduplication using a \textsc{HashSet} to track recently collected results, and (2) a 4-second throttling interval to prevent duplicate saves. To allow re-collection of legitimately reoccurring content (e.g., when users scroll back to previously viewed screens), we clear the deduplication cache every 10 seconds. As a result, this strategy balances collection efficiency with completeness.
 
 The code for \sys{} will be open-sourced to enable easy adoption and community-supported improvements.}

\textcolor{black}{\subsection{Open Sourcing and Deployment Options}}

\textcolor{black}{\sys{} will be fully open-sourced, offering complete customizability and self-hosting options for specific research and regulatory needs. While our reference implementation uses Firebase Realtime Database, the data collection architecture is built on a modular API layer that can be adapted to work with any backend infrastructure. Researchers can deploy \sys{} with self-hosted databases (e.g., MongoDB, PostgreSQL\footnote{MongoDB: https://www.mongodb.com, PostgreSQL: https://www.postgresql.org}) or cloud services that meet their institutional requirements.}

\textcolor{black}{The ability for everyone to self-host \sys{} is particularly important for studies with sensitive data or regulatory constraints, as different countries and territories have specific requirements about data storage location and cross-border data transfer. Standard Firebase may not meet certain compliance requirements (e.g., FERPA, HIPAA, GDPR), but enterprise configurations or alternative infrastructures can address these needs. We encourage researchers to consult with their institutional review boards and IT security teams when deploying \sys{} to ensure compliance with relevant data protection regulations.}

\section{Evaluations}

To test and validate \sys{}, we conducted a series of three evaluation studies. The first two studies focused on researchers' experience to collect data (Study 1) and \sys{}'s data collection accuracy (Study 2). The third study investigated \sys{}'s battery usage and perceived interruptions in the wild for participants (i.e. data contributors) (Study 3). The first two studies were designed to be in-lab, while Study 3 was conducted as a field study. These three studies together aim to answer the following research questions:

\begin{enumerate}
\item RQ1: How well does \sys{} support researchers in collecting mobile screen data (Study 1)?
\item RQ2: How accurately does \sys{} collect target data on participants' screens (Study 2)?
\item RQ3: How does \sys{} influence device performance for participants (i.e. data contributors) in a real-world data collection scenario (Study 3)?
\end{enumerate}

\subsection{Evaluation Scenarios}
\label{sec:usage-scenarios}
We wanted our performance evaluation of \sys{} to cover a range of different data collection needs. \textcolor{black}{Looking broadly into previous research in HCI, CSCW, and Data Science, we extracted three scenarios where researchers can use \sys{} to collect user screen data. Our evaluation studies were based in these scenarios.}

\begin{figure*}
    \centering
    \includegraphics[width=\linewidth]{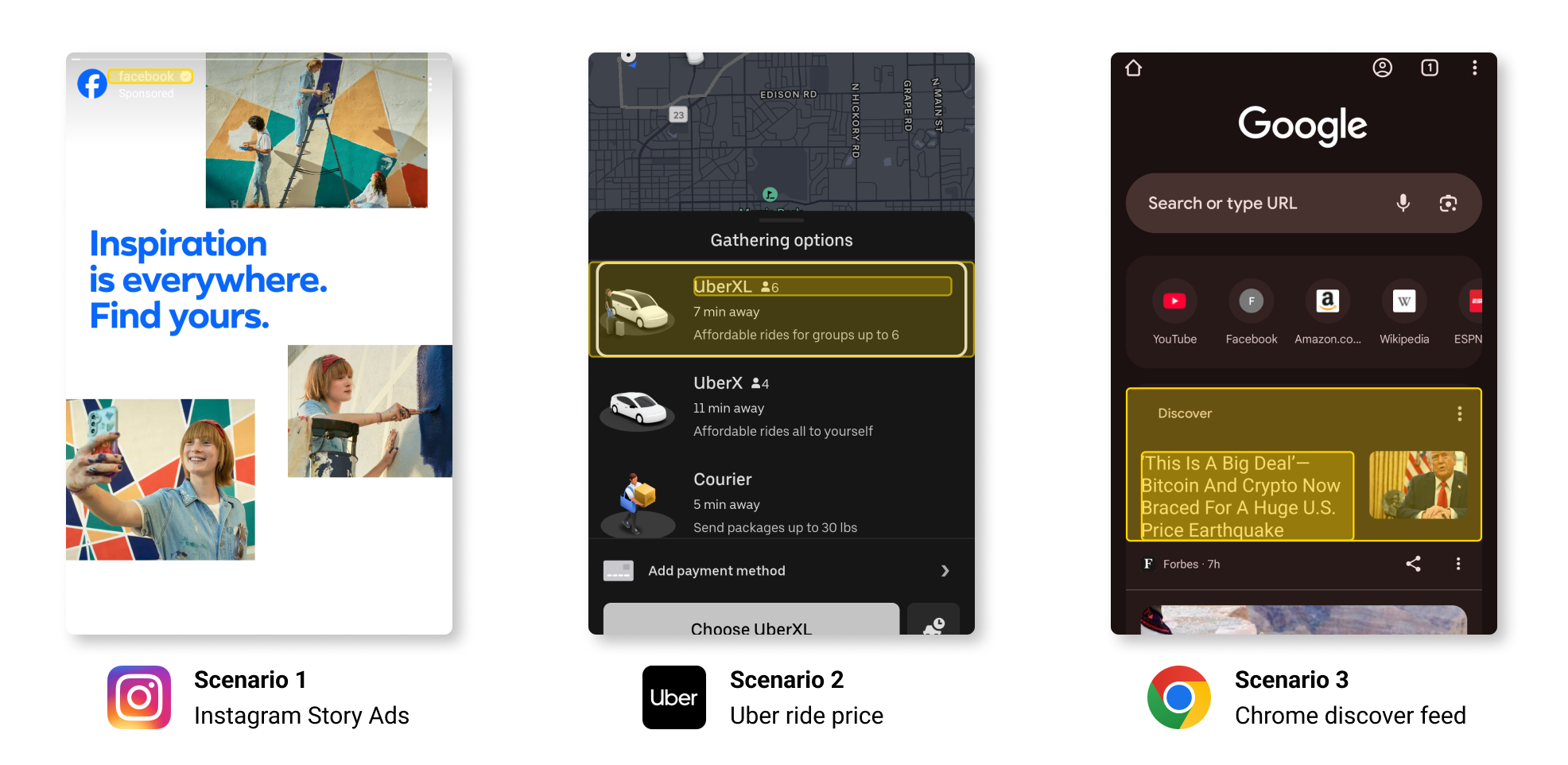}
    \caption{We identified three potential usage scenarios of \sys{} for our evaluation. Scenario 1 collects the advertiser information in Instagram Story ads, scenario 2 collects Uber ride pricing when user requests a ride, and scenario 3 collects the top feed item in Chrome Discover Feed recommendations. \sys{} shows a yellow highlight overlay every time the target data is collected, providing full transparency to the collection participants.}
    \label{fig:eval_scenarios}
    \Description{This figure shows three mobile phone screenshots arranged horizontally, each labeled as a different scenario. In Scenario 1 (left), an Instagram Story advertisement is shown with the text ``Inspiration is everywhere. Find yours.'' along with several colorful images featuring someone in casual settings. In Scenario 2 (middle), an Uber app interface displays ride options near a Walmart Supercenter. A yellow highlight box surrounds "UberXL" pricing information showing "7 min away" and "Affordable rides for groups up to 6." In Scenario 3 (right), a Google Chrome mobile browser interface shows the Discover feed. A yellow highlight box surrounds a news article preview titled ``This Is A Big Deal - Bitcoin And Crypto Now Braced For A Huge U.S. Price Earthquake'' from Forbes published 7 hours ago. Each screenshot represents a different type of digital data that could be collected for research purposes, with the yellow highlighting indicating the specific elements of interest.}
\end{figure*}

\paragraph{Scenario 1: Instagram Story Advertisements}
Instagram Story advertisements are sponsored short image/video segments that appear between an Instagram user's following content. They usually come with a ``Sponsored'' label to be distinguished from users' regular following content. Mutiple research studies have been conducted in recent years around Instagram Story advertisements, to understand how users perceive and understand this specific form of advertising~\cite{balaban_adolescents_2022, haldborg_jorgensen_instagram_2023, fakhradyan_study_2021}. However, it is not easy to collect a dataset of Instagram story advertisements.  

Getting access to a dataset of Instagram's Story ads can enable research contributions in fields including social media research~\cite{hanbazazh_pop-up_2021}, algorithm auditing~\cite{ulbricht_algorithmic_2022, eslami_i_2015, eslami_first_2016}, and usable privacy~\cite{chen_empathy-based_2024}. Specifically, we can empirically observe users' interactions with Instagram Story ads, as well as inferring dwelling time over Story ads based on the first and last content change while a specific ad appears on screen. More generally, some broader questions we can answer with the dataset include: What empirical understanding of Instagram's personalized advertisement algorithms can we obtain from quantitative data analysis? How do Instagram Story advertisements reveal the platform's portraits of individual users' preferences? And what are users' folk theories of Instagram's personalized ad algorithm?\looseness=-1

\paragraph{Scenario 2: Uber Pricing}
Ride-sharing platforms like Uber implement dynamic pricing strategies that adjust fare prices based on various factors including demand, time of day, and location. Collecting real-time price data for trip requests would allow researchers to study pricing algorithms, examine fare consistency across different user groups, and analyze how external factors influence ride-sharing costs. Previous research has attempted to understand these pricing mechanisms through qualitative measure~\cite{lenjo2017qualitative} and quantitative analysis~\cite{chao2019modeling}. However, current data sources cannot cover many user-specific aspects such as price variations across user groups and discount information. This data could help researchers investigate questions about algorithmic fairness in transportation access, analyze the relationship between surge pricing and local events, and study how pricing strategies affect user behavior in different geographical and temporal contexts. In our scenario, we focus on collecting the pricing information for participants' Uber ride requests.

\paragraph{Scenario 3: Chrome Browser Discover Feed}
Mobile browsers like Google Chrome integrate algorithmic news feeds into their search interface. These algorithmic feeds, often supported by deep learning, curates content based on users' browsing behavior and inferred interests~\cite{zheng2018drn}. These algorithmic feeds, while work at an individual user level, have wide and profound social implications~\cite{sandvig2014auditing, ribeiro2020auditing}.

Using Chrome as an example, accessing its Discover feed data from the user's perspective enables research in algorithm auditing, helping to analyze content recommendation patterns, assess personalization mechanisms, and identify potential biases~\cite{sandvig2014auditing, guess2023social}. Researchers have proposed audit studies to uncover discriminatory practices~\cite{sandvig2014auditing}. Researchers can investigate how these algorithms shape information exposure by examining content diversity, frequency, and source representation. Understanding these dynamics in the Discover feed can identify biases impacting users' information access and inform design improvements for more equitable content curation.

\subsection{Study 1: Evaluating \sys{} Usability with Researchers (In-Lab)}
\label{sec:eval-one}

Study 1 evaluates how \sys{} effectively supports \textbf{researchers}. While \sys{} serves both researchers (collector creators) and users (data contributors), this study focuses specifically on the researcher experience during collector creation. We examine the usability of our Programming By Demonstration (PBD) approach in creating a data collector and try to understand what are researchers' potential use cases of \sys{}.

\subsubsection{Procedure}
Each one-hour session began with a brief introduction to \sys{}'s motivation and core features. Participants then created data collectors for three representative usage scenarios (detailed in \cref{sec:usage-scenarios}) using \sys{}'s interface. After completing the tasks, participants completed a modified System Usability Scale (SUS) questionnaire~\cite{brooke1996sus}, including five-point Likert scale questions evaluating \sys{} on its usability and utility for future research projects. We concluded with semi-structured interviews exploring participants' envisioned uses of \sys{} in their research. All studies were carried out in person and participants received a \$15 digital gift card for their time. To evaluate the accuracy of the PBD-generated Graph Queries, two authors independently analyzed whether each Query correctly captured the researcher's specified data collection targets.

\subsubsection{Participants}
We recruited 5 HCI researchers through social media advertisements and word of mouth. All participants had experience with data collection and had interests in collecting users' screen data for research purposes. Most participants have experience collecting various types of user data in the past (see participant demographics in \cref{appendix-eval-one-demo}). 


\begin{figure*}
    \centering
    \includegraphics[width=\linewidth]{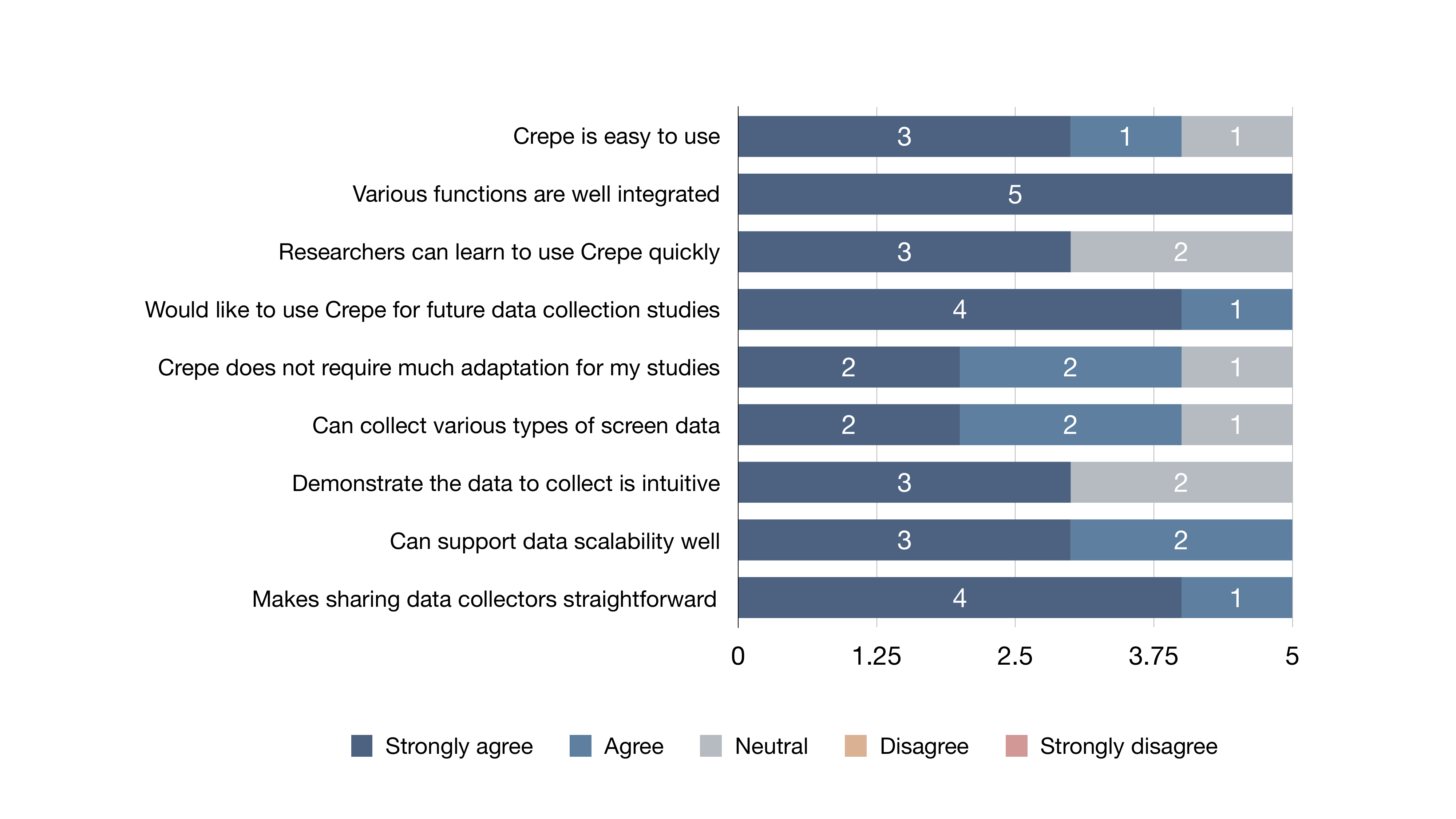}
    \caption{The usability questionnaire results for Study 1, evaluating researchers' perception of Crepe's usability and data collection capabilities. The tool received positive feedback, demonstrating its usability and utility for potential data collections studies.}
    \label{fig:study_1_results}
    \Description{A horizontal bar chart showing responses to 9 statements about Crepe's usability on a 5-point Likert scale from 'Strongly agree' to 'Strongly disagree'. Most responses trend positive, with 'Various functions are well integrated' receiving the strongest agreement (5 respondents strongly agreeing). The chart uses a color scheme of dark blue for 'Strongly agree' through pink for 'Strongly disagree', with statement text aligned left and response counts displayed in bars.}
\end{figure*}

\subsubsection{Results}
The usability evaluation showed positive results (\cref{fig:study_1_results}), with participants rating \sys{} favorably on all questions. Most participants also showed a strong interest in using \sys{} for future screen data collection studies and think it would not require much adaptation to \sys{}. 

In Study 1, as researchers created data collectors using \sys{}, we also evaluated the programming by demonstration (PBD) pipeline for authoring candidate Graph Queries. Our evaluation was carried out on in total 15 created collectors (3 example scenarios for each of the 5 participants). We found that in \textit{all of} the 15 occasions (15/15, 100\%), our PBD pipeline successfully generated Graph Queries that correctly match users' data collection goal (see \cref{fig:query_details}). Based on participant feedback, we iteratively improved the design of the user disambiguation process. We landed on our final design as described in \cref{sec:user-intent-disambiguation}, making it intuitive to pick the best Graph Query that matches researchers' data collection goals. 


Our participants identified several novel research opportunities enabled by \sys{}'s screen data collection capabilities. P1 mentioned the interest in collecting Twitter (X)'s feed data to understand the platform's feed algorithm. P2 would like to collect data from Uber to understand how ride prices change throughout a day and throughout different days of a week. P5 proposed to collect journal apps' suggestion prompts to gauge what the app understands about users' everyday activities. \textcolor{black}{Interestingly, P1 also brought up the need to collect data based on certain conditions: for example, only collect social media feed under the ``For you'' tab instead of ``Following''. This can further expand the utility of \sys{}.}\looseness=-1

\subsection{Study 2: Evaluating Graph Query Data Collection Accuracy (In-Lab)}
\label{sec:eval-study-two}
Study 2 evaluates how accurately \sys{} collects screen data in realistic app usage scenarios. While Study 1 focused on researchers' experience creating data collectors, this study examines the technical performance of Graph Query in locating and extracting target data.\looseness=-1

\textcolor{black}{For mobile screen data collection tasks, it is not easy to establish accurate ground truth, i.e., all of the data we are supposed to collect. Measuring screen data collection accuracy requires us to compare (1) the target data that showed up on users' screens (i.e., the ground truth), and (2) the data the tool is able to collect. It is almost impractical or impossible to establish the ground truth when the data collection happens in the wild, as we need to screen record all user interactions and manually identify target data on screen. To the best of our knowledge, no automated tool is capable of identifying target data with 100\% accuracy, especially given \sys{}'s specificity towards target data. A previous tool that collects all screen data during a period was not able to compare collected screen data against a rigorous ground truth~\cite{teng2024tool}. Instead, they indirectly proved their collected data contained ``no notable anomalies'' and they generally followed their expected usage patterns. As a result, to accurately evaluate \sys{}'s data collection accuracy, we conducted it as an in-lab study, so we can record the screen and manually label a ground truth dataset.}

\begin{table*}[htbp]
\newcolumntype{Y}{>{\centering\arraybackslash}X}
\begin{tabularx}{\textwidth}{lYYYY}
\toprule
\textbf{Metric} & \textbf{Overall} & \textbf{Instagram} & \textbf{Uber} & \textbf{Chrome} \\
\midrule
\textbf{Recall}    & 95.80\% (506/528) & 94.7\% (269/284)  & 96.6\% (115/119)  & 97.6\% (122/125) \\
\textbf{Precision} & 96.00\% (504/525) & 95.0\% (268/282)  & 100\% (115/115)   & 94.5\% (121/128) \\
\textbf{F1 Score}  & 96.0\%           & 94.8\%           & 98.3\%           & 96.0\%         \\
\bottomrule
\end{tabularx}
\caption{The evaluation results for \sys{} in the three app scenarios. Our in-lab study 2 demonstrates that \sys{} performs well on on the three selected data collection scenarios. We discuss the observed edge cases that \sys{} cannot handle well in \cref{sec:edge-cases}.}
\Description{This table presents performance metrics for a classification system across three digital platforms (Instagram, Uber, and Chrome) along with overall performance measures, where the total number of entries analyzed varies by platform with 528 entries overall, 284 entries for Instagram, 119 entries for Uber, and 125 entries for Chrome. The recall (true positive rate) measurements show 506 out of 528 (95.80\%) overall, 269 out of 284 (94.7\%) for Instagram, 115 out of 119 (96.6\%) for Uber, and 122 out of 125 (97.6\%) for Chrome. The precision (positive predictive value) results indicate 504 out of 525 (96.00\%) overall, 268 out of 282 (95.0\%) for Instagram, 115 out of 115 (100\%) for Uber, and 121 out of 128 (94.5\%) for Chrome. The F1 scores (harmonic mean of precision and recall) are 96.0\% overall, 94.8\% for Instagram, 98.3\% for Uber, and 96.0\% for Chrome, with Uber achieving perfect precision but slightly lower recall, while Chrome demonstrated the highest recall but the lowest precision, and Instagram maintained consistent performance across all metrics around 95\%, resulting in strong overall performance across all platforms with F1 scores above 94\%.}
\label{tab:eval-results}
\end{table*}

\subsubsection{Procedure}
We provided participants with a development phone with \sys{} pre-installed and configured to collect data from apps in the three defined scenarios (\cref{sec:usage-scenarios}). Each participant spent approximately one hour using these apps, allocating around 20 minutes per app. Participants were instructed to use each app naturally as they would in their daily life, while also feeling free to switch between apps when needed to mirror realistic usage patterns. All studies were carried out in person and participants received \$15 digital gift cards for each hour they participated.

To establish ground truth for evaluating collection accuracy, we simultaneously recorded the phone screen during the study sessions. An author labeled the screen recordings to identify all instances where target data appeared as the ground truth, and another author separately verified the results. We compared this ground truth data against the data collected by \sys{} to compute collection accuracy.

\subsubsection{Participants}
We recruited 5 participants through social media advertisements and word-of-mouth. This provided approximately 240 minutes of naturalistic app usage data for our accuracy evaluation, generating a diverse set of UI states and interactions across different apps. Participant demographics and their app usage experience are detailed in \cref{appendix-eval-two-demo}.

\subsubsection{Results}
Across all sessions, we recorded 528 data points across three apps as the benchmark (284 from Instagram, 119 from Uber, and 125 from Chrome). \sys{} achieved an overall collection accuracy F1 score of 96.0\% (precision: 96.0\%, recall: 95.8\%). Breaking down by app, collection accuracy F-1 scores were consistently high: Uber (98.3\%), Instagram (94.8\%), Chrome (96.0\%). The minor variations in accuracy across apps can be attributed to differences in UI update frequencies and layout complexities.

{\color{black}
\label{sec:edge-cases}
\paragraph{Edge cases.} In the evaluation, we observed several scenarios where \sys{} can make mistakes in data collection. First, some apps have UI layout variations for the same target data, requiring \sys{} to set up multiple Graph Queries to collect the same data (as discussed in \cref{sec:query-limitations}). In our evaluation, Instagram story's ads are sometimes presented above a \textbf{TextView} containing ``Sponsored'', and sometimes above a \textbf{Button} containing ``Sponsored''. However, by demonstrating in both of these scenarios, researchers can still use \sys{} to robustly handle such data collection tasks. In fact, we demonstrated three different UI layouts to \sys{} during our evaluation study for Instagram Stories. Second, complex UI screens containing more than 500 UI elements can pose challenges to \sys{}'s processing speed, potentially omitting data due to performance bottlenecks. We set a loose threshold to \sys{}'s processing frequency to avoid repeatedly collecting the same data. In addition, certain techniques used in App development, such as virtualization, might cause Android Accessibility Service to pick up invisible text on screen. This is also observed in a previous study~\cite{teng2024tool}. We discuss some of our future improvement plans to address these challenges in \cref{sec:future-work}.
}

In general, the results show that Graph Query can reliably collect screen data across various apps and UI screens. The collected dataset captured diverse scenarios including real-time price updates in Uber, dynamically positioned sponsored content in Instagram, and recommendation feeds in Chrome homepage. These results validate that Graph Query's data collection mechanism is robust enough for real-world deployment while maintaining high accuracy across different apps and scenarios. 

\subsection{Study 3: Evaluating System Impact for Users (Field Study)}

Now that we understand \sys{}'s performance in lab, how well does it work in the wild? In Study 3, we conducted a field study with users who contributes their screen data through \sys{}. Study 3 focuses on the real-world performance of \sys{}, especially on its device performance and impact on data contributors' regular phone usage experience. We designed Study 3 to evaluate how \sys{} performs on different Android devices and OS versions, helping us catch any real-world deployment issues we couldn't spot in the lab. Study 3 was conducted in example scenario 1 (Instagram story ads), since our target metrics will not significantly change when deployed for different apps' data collections.

\begin{table*}[!htbp]
\small
\begin{tabularx}{\textwidth}{>{\hsize=0.7\hsize}X>{\hsize=1.6\hsize}X>{\hsize=1.2\hsize}X>{\hsize=0.6\hsize}X>{\hsize=0.9\hsize}X}

\toprule
\textbf{Participant} & \textbf{Mobile OS} & \textbf{Device} & \textbf{Period} & \textbf{Battery Usage} \\
\midrule
PC1 & Android 14 & Pixel 8 (own) & 72 hours & 1\% \\

PC2 & Android 14 & Pixel 6 (provided) & 72 hours & 14\% \\

PC3 & Android 14 & Pixel 6 (provided) & 72 hours & 9\% \\

PC4 & Android 13, OxygenOS 13.1 & OnePlus 9R (own) & 24 hours & 0.91\% \\

PC5 & Android 12, Oxygen OS 12.1 & OnePlus Nord (own) & 24 hours & N/A \\

PC6 & Android 14 & Pixel 6 (provided) & 24 hours & 2\% \\

PC7 & Android 9 & Galaxy S10 (own) & 24 hours & 0\% \\
\bottomrule
\end{tabularx}
\caption{Mobile device usage data for study participants over 24 to 72 hour periods, including mobile OS, device model, and battery usage percentage. Three of the participants did not have an Android phone and used our provided device, with which they were asked to use Instagram following their regular habits during the study period. The participants decided the time period of the study based on their time schedule. Note that P5 reported that they could not find the battery usage item of \sys{} on their phone Settings page after the collection. We were not able to help them find it either, despite our best efforts.}
\Description{This table presents mobile device usage data for seven participants over study periods ranging from 24 to 72 hours. The table includes each participant’s mobile OS, device model, study duration, and the battery usage attributed to running the Crepe app. Three participants (PC2, PC3, and PC6) used the provided devices (Pixel 6 with Android 14) because they did not own an Android phone. The remaining participants used their own devices, which varied in model and Android version. Battery usage varied by participant and device. PC1 used a Pixel 8 for 72 hours and reported 1\% battery usage. PC2 and PC3, both using provided Pixel 6 devices for 72 hours, reported 14\% and 9\% usage, respectively. PC4 used a OnePlus 9R (Android 13, OxygenOS 13.1) for 24 hours and reported 0.91\% battery usage. PC5, using a OnePlus Nord (Android 12, OxygenOS 12.1) for 24 hours, was unable to locate battery usage data in their device settings. PC6 used a provided Pixel 6 for 24 hours and reported 2\% usage. PC7 used a Galaxy S10 (Android 9) for 24 hours and reported 0\% battery usage. The results suggest that Crepe consumes minimal battery across a range of Android devices and OS versions. Most participants reported battery usage under 2\%, indicating that Crepe is energy-efficient for background data collection. The only exception was PC2, who experienced 14\% usage over 72 hours. A note below the table explains that PC5 could not find the battery usage stats despite assistance from the research team.}
\label{mobile_usage_data}
\end{table*}

\paragraph{Participants}
We recruited seven participants through social media advertisements and word-of-mouth. The participants are regular users of Instagram and Instagram Story. Each participant decided to contribute their data from 24 hours up to 72 hours based on their preferences and time schedule. We aimed at testing \sys{} on various Android versions and device platforms, while also evaluating \sys{}'s performance over short and longer collection durations. Four participants used their own Android devices. Since our main goal is not to collect a dataset that can be directly used in rigorous quantitative analysis but to empirically test our collector system, we also recruited three participants who are interested but do not have an Android device (P2, P3, P6). We provided each of them with an Android phone, with which they were asked to use Instagram during the period of the study, roughly following their regular Instagram usage habits. These three participants also contributed to our study by helping us assess the performance and usability of \sys{} across different usage behaviors.

\paragraph{Study procedure}
The user study followed a three-stage procedure\footnote{The study protocol was reviewed and approved by the Institutional Review Board (IRB) at our institution.}:

\begin{enumerate}
\item A 30-minute introductory interview to familiarize participants with the study and set up \sys{} on their devices;
\item A testing period lasting between 24 to 72 hours, based on the participants' time schedule and preferences, during which they were encouraged to use Instagram as usual.
\item A 30-minute concluding interview to administer a post-study questionnaire, collect feedback on participants' experiences, and address any questions or concerns.
\end{enumerate}

Throughout the study, we maintained email communications with participants to ensure \sys{} functioned normally and conducted troubleshooting sessions when necessary. Each participant was compensated with 25 USD digital gift card for the data they contributed, and 20 USD for each hour of meeting sessions they attended. The participants' participation duration, device information, and \sys{}'s battery usage result after the study ended are shown in Table \ref{mobile_usage_data}.

\begin{figure*}
    \centering
    \includegraphics[width=\linewidth]{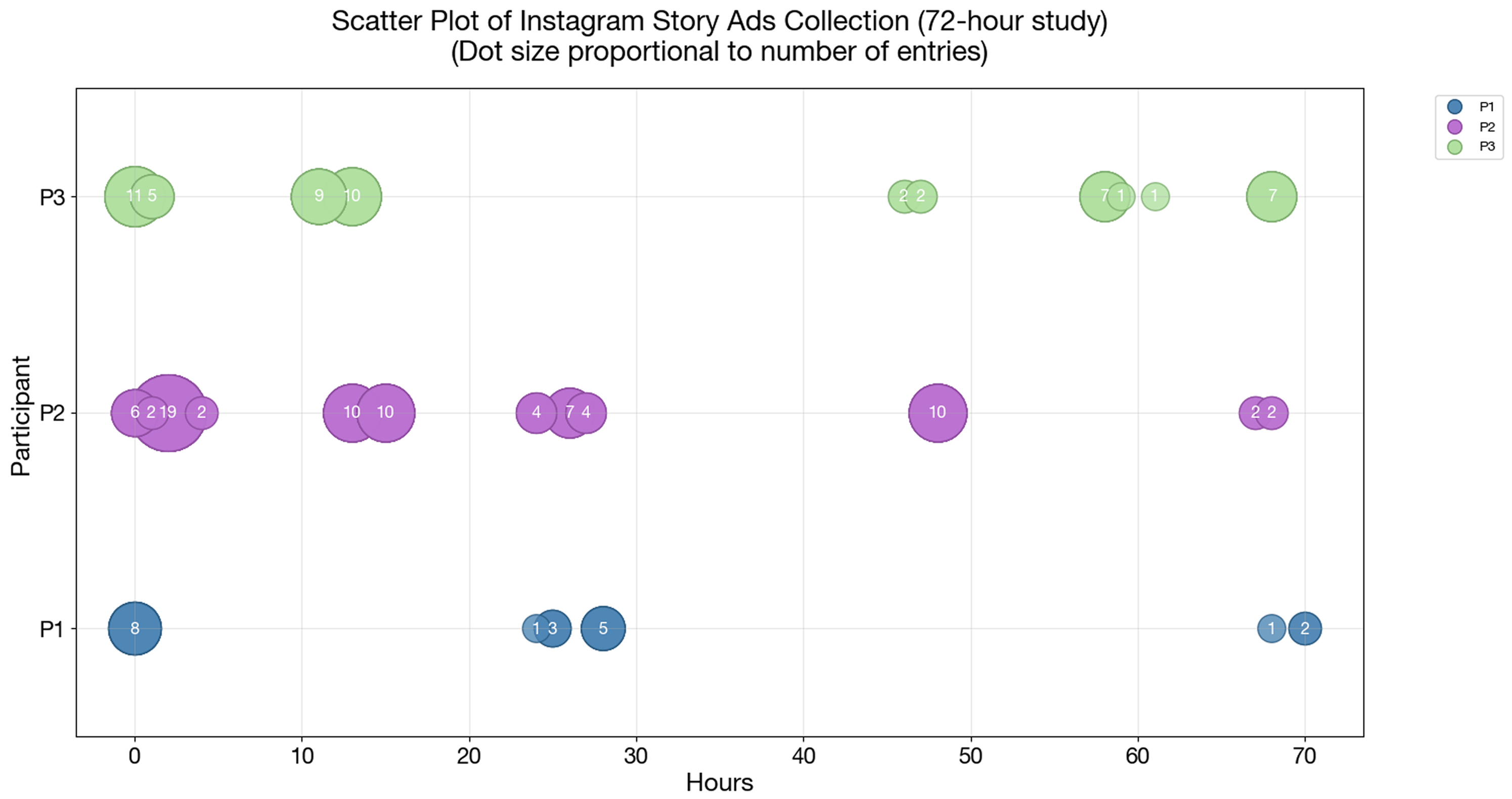}
    \caption{Temporal patterns of Instagram story ads collected in Study 3 using \sys{}. This scatter plot displays unique data from participants P1--P3, who chose a 72-hour study period based on their availability. Circle size corresponds to the number of entries collected at each time point (shown by the number within each circle). The distribution reveals how ad exposure varied across participants and throughout the three-day period, with some participants experiencing more consistent ad delivery while others encountered more sporadic patterns. All data was passively collected while participants engaged in their regular smartphone usage. \textcolor{black}{For the rest of the participants, the 24-hour study results is shown in \cref{fig:study3_1day}.}}
    \label{fig:study3_3day}
    \Description{This image shows a scatter plot titled ``Scatter Plot of Instagram Story Ads Collection (72-hour study)'' with the subtitle ``Dot size proportional to number of entries.'' The plot displays data for three participants (P1, P2, and P3) over a 72-hour period. The y-axis represents the participants, labeled from P1 at the bottom to P3 at the top. The x-axis shows time in hours, ranging from 0 to approximately 70 hours. Each data point is represented by a colored circle where the size of the circle corresponds to the number of Instagram story ads collected, with larger circles indicating more entries. P1 (shown in blue circles) has data points scattered at hours 0 (8 entries), around hour 25 (1, 3, and 5 entries), and at hour 70 (1 and 2 entries). P2 (shown in purple circles) has more frequent data collection with clusters at hours 0-5 (with entries of 6, 2, 19, and 2), hours 10-15 (with 10 entries multiple times), hours 20-30 (with 4, 7, and 4 entries), hour 45 (with 10 entries), and hour 65 (with 2 entries twice). P3 (shown in green circles) has data points at hours 0-5 (15 entries), hours 10-15 (9 and 10 entries), hours 40-45 (2 entries twice), hours 55-60 (7, 1, and 1 entries), and hour 65 (7 entries). The legend in the top right identifies each participant with their corresponding color: P1 (blue), P2 (purple), and P3 (green).}
\end{figure*}

\begin{figure*}
    \centering
    \includegraphics[width=\linewidth]{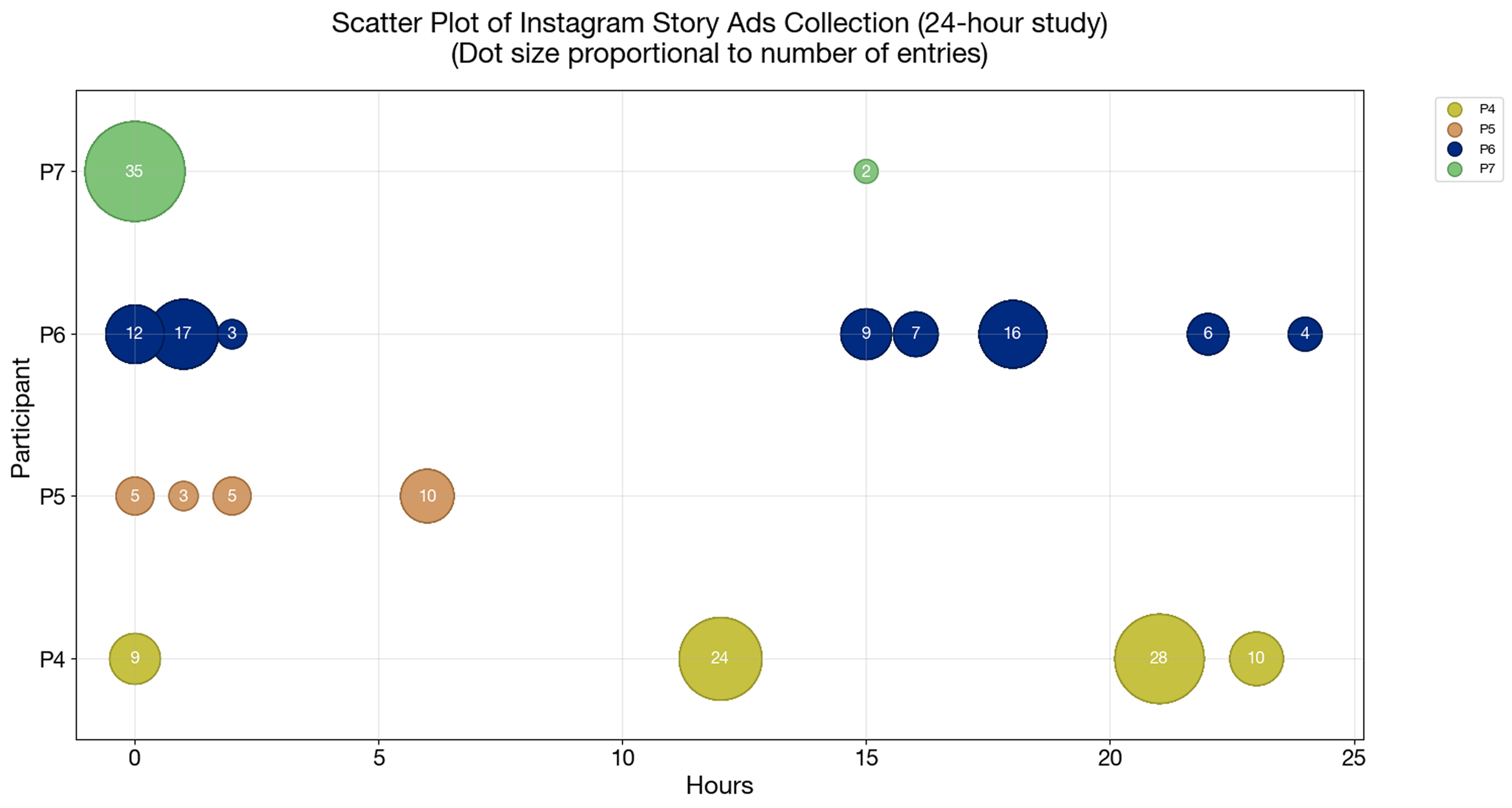}
    \caption{The distribution of Instagram story ads collected in Study 3 using \sys{}. This scatter plot represents unique data collected from participants P4-P7, who chose a 24-hour study period based on their availability. Each circle represents a collection event, with size proportional to the number of ads captured (indicated by the number inside each circle). Participants installed \sys{} on their mobile phones and continued their normal daily activities while the system collected Instagram story ads in the background.}
    \label{fig:study3_1day}
    \Description{This image displays a scatter plot titled ``Scatter Plot of Instagram Story Ads Collection (24-hour study)'' with the subtitle ``Dot size proportional to number of entries.'' The plot tracks data collected from four participants (P4, P5, P6, and P7) over a 24-hour period. The horizontal axis represents time in hours (from 0 to 25), and the vertical axis shows the four participants. Each data point appears as a colored circle — olive for P4, orange-brown for P5, dark blue for P6, and light green for P7. The size of each circle corresponds to the number of Instagram story ads collected at that time point, with larger circles representing more entries. The specific number of entries is displayed within each circle. Participant P7 (light green circles) has the largest single collection of 35 entries at hour 0 and a smaller collection of 2 entries around hour 15. Participant P6 (dark blue circles) shows consistent activity throughout the 24-hour period, with multiple collection points at hours 0-2, 15-16, 18, 20, and 23. Participant P5 (orange-brown circles) shows activity primarily in the first 5 hours, with entries of 3-10 ads collected. Participant P4 (olive circles) shows a pattern of data collection at the beginning (hour 0), middle (hour 10), and end (hours 18-22) of the study period, with the largest collections of 24 and 28 entries occurring in the latter half of the study.}
\end{figure*}

\paragraph{Study Results}
After initial data analysis, \sys{} was able to collect 358 Instagram Story advertisements from all seven participants, with an average of 51 ads per participant (\cref{fig:study3_3day}, \cref{fig:study3_1day}). The collected data reveals the number of unique Instagram Story Ads encountered by each participant at regular time intervals throughout their respective data collection periods. The data allow us to observe the frequency and distribution of Instagram Story Ads over time, potentially providing insights into the advertising strategies employed by businesses on the platform. We shared the data we collected from each participant with them individually upon request.

In the post-study questionnaire, which included two 5-scale questions, all participants unanimously reported that they felt \textbf{minimal interruption and disturbance} from \sys{} while using their everyday apps and \textbf{did not notice significant changes} to their devices during the collection study. Here are details regarding these two questions and all participants' response:

\begin{itemize}
    \item Q1: When using my everyday apps, I felt minimal interruption and disturbance from the Crepe app.
    \begin{itemize}
        \item All participants responded 5 – Strongly agree
    \end{itemize}
    \item Q2: I noticed significant changes to my device during the data collection study.
    \begin{itemize}
        \item All participants responded 1 – Strongly disagree
    \end{itemize}
\end{itemize}

Based on the battery usage data collected in the post-study questionnaire, Crepe demonstrated excellent energy efficiency on participants' mobile devices over the 24 to 72 hour study periods. As shown in Table \ref{mobile_usage_data}, the battery usage of Crepe ranged from 0\%, 0.91\% to 14\%, with most participants reporting usage of 2\% or less. This indicates that running Crepe in the background for data collection has minimal impact on the battery life of participants' smartphones, making it a power-efficient tool for mobile data collection studies.

However, we observed a limitation of \sys{} during the study: the background process management on Android can terminate \sys{} to conserve resources. This is a known issue for similar mobile data collectors~\cite{ferreira_aware_2015}. For P1, we suspect the Android operating system terminated \sys{} in the background, resulting in data loss at around hour 30. Only after we noticed it and prompted P1 to check the running status of \sys{} did the data resumed to be collected. To address this issue, for later participants, we improved \sys{}'s ``heartbeat'' mechanism and had the app send a heartbeat signal every 10 minutes. Once we noticed a potential inactive status of \sys{}, we sent reminder emails to participants and asked them to open \sys{} and check its running status. In future iterations of \sys{}, this heartbeat mechanism could be combined with a dashboard to alert researchers when data collection becomes offline, enabling timely interventions to minimize data loss. \looseness=-1
\subsection{Summary}

Our three-part evaluation of Crepe examined its effectiveness in supporting researchers' data collection needs (RQ1), accuracy in collecting target screen data (RQ2), and real-world performance impact (RQ3). Study 1 demonstrated that researchers could effectively use Crepe's programming by demonstration approach, with all five participants successfully creating accurate data collectors for three different scenarios and rating the system's usability favorably. Study 2 validated Crepe's technical robustness through an in-lab evaluation, achieving high collection accuracy with an overall F1 score of 96.0\% across Instagram, Uber, and Chrome applications, though some challenges emerged with complex UI screens containing over 500 elements. Finally, Study 3's field deployment with seven participants showed that Crepe had minimal impact on device performance, with most participants reporting battery usage of 2\% or less and unanimously indicating no noticeable interference with their regular phone usage. The system successfully collected 358 Instagram Story advertisements during the field study, though we identified opportunities for improving background process management to prevent occasional data collection interruptions. Together, these results demonstrate that Crepe effectively balances researcher needs, technical accuracy, and user experience in mobile screen data collection, while highlighting specific areas for future enhancement.
\section{Discussion}

\subsection{Collection Support Tools for De-centralized Data Access}
\sys{} complements existing open-source data collection tools like AWARE~\cite{ferreira_aware_2015} and Purple~\cite{schueller_purple_2014} that focus on collecting sensor data, and others like ShiptCalculator~\cite{calacci_bargaining_2022} that enable workers to share screenshots of their pay history for wage transparency advocacy. \sys{} targets the generalized collection of various types of mobile screen data. This contains great potential for more academic researchers to more easily access datasets and uncover more quantitative, analytical insights. \sys{} aligns with the growing movement towards a more open and equitable ecosystem in digital economy~\cite{arrieta-ibarra_should_2018}.\looseness=-1

The democratization of screen information access breaks the ``data monopoly'' that operating systems providers and app developers have long maintained~\cite{jia-jun_li_bottom-up_2022}. It shifts the data ownership back to end users, instead of leaving it only concentrated in the hands of tech platforms. By helping users see their own data, we can enable customized features that monitor and understand their behavior, supporting more informed device and app usage. This transparency allows individuals to make better decisions about their digital habits and identify patterns they might want to change.

By providing academic researchers access to screen data, they can better audit algorithms without depending on platform APIs that companies can (and often do) remove at will~\cite{ribeiro2020auditing, sandvig2014auditing, calacci_bargaining_2022, kaplan2024comprehensively}. This independent access creates a more sustainable research ecosystem that isn't vulnerable to corporate gatekeeping. Researchers can investigate questions about algorithmic bias, content recommendation systems, or user experience issues with greater freedom and thoroughness. This approach represents a fundamental shift in how we think about digital data ownership, treating screen data as something that belongs primarily to users rather than platforms. It encourages a more balanced relationship between tech companies, users, and the research community.

{\color{black}
\subsection{Graph Query for Post-Collection Data Filtering}
While \sys{}'s current design focuses on prospective data collection where researchers define Graph Queries before deployment, a potential alternative approach is to collect comprehensive app-level data first similar to existing tools~\cite{teng2024tool}, then use Graph Query retrospectively to filter relevant information. This provides the benefit of retaining contextual information about the full UI state and enabling researchers to iteratively refine their data collection criteria without redeploying to participants. In this way, researchers could iteratively refine Graph Queries on stored UI snapshots, similar to writing SQL queries to explore structured data without re-collecting, enabling exploratory analysis and hypothesis refinement post-deployment. But at the same time, this retrospective approach would require the initial data collection phase to store UI Snapshot information (including all screen entity relations) rather than only the final extracted values. This comprehensive collection increases storage requirements and privacy concerns but provides flexibility for researchers to discover unexpected patterns or change research questions post-collection, with Graph Query serving as the unified querying mechanism for both use cases.
}

\subsection{Careful Considerations for User Privacy}
The collection of user data through \sys{} raises important privacy concerns regarding users' app usage and screen information. Throughout the design and implementation of \sys{}, user privacy has been a top priority, ensuring real-time transparency of collected data and employing high-standard encryption for data security. It is crucial to emphasize that \sys{} is intended solely for research purposes, under scenarios where researchers obtain explicit consent from participants to collect screen data. In the future, we plan to continuously enhance the privacy-related features in \sys{}. For instance, prior to the public release of \sys{}, we intend to include a feature that hashes personal identifiers before sending them to researchers, thereby protecting users' privacy, similar to the approach taken by previous data collectors~\cite{ferreira_aware_2015}. There have also been differential-privacy-based approaches that obfuscate sensitive data on collected mobile app UI snapshots~\cite{li2020privacy}, which we can adopt in \sys{}. We acknowledge that protecting users' privacy in data collection remains an open research question, and we encourage the research community to collectively work towards improving our ethical standards in this regard.

\subsection{Implications for Data Labor and Data Governance}
Our approach for the decentralization of data access has important implications for data governance and the broader ecosystem of data-driven technologies. The rise of digital platforms has led to the concentration of user data in the hands of corporations and platforms, creating a \textit{``data monopoly''} that limits the potential benefits of academic research and user-centric data-driven technollgies~\cite{jia-jun_li_bottom-up_2022}. By breaking the data monopoly and providing users with access to their own data, our approach can help bridge the data divide and empowers individuals and communities to leverage their own data for personal and collective benefits, rather than being solely exploited by corporations.

This increased transparency can help to identify and address biases and inequalities in the algorithms~\cite{panch_artificial_2021, kordzadeh_algorithmic_2022}. By providing access to users' own data on platforms, distributed, user-centric data-driven technologies can be developed in user groups and communities to shift the focus of benefits from corporations and platforms to end users. This aligns with the vision of a more equitable data ecosystem, where the value generated by user data is distributed more fairly among stakeholders~\cite{posner_radical_2018}.

{
\color{black}

\section{Limitations and Future Work}
\label{sec:future-work}

While \sys{} offers a novel approach to mobile screen data collection, we acknowledge several limitations and opportunities for future enhancement.

\subsection{Evaluation Study Scale}

Our evaluation consists of three segmented studies, each designed to assess specific aspects of \sys{} with practical constraints inherent to each study type. While these studies validate core functionality and usability, they involve smaller participant pools and controlled scenarios. Our Study 2 was also limited by the fact that it was conducted using one device type. We acknowledge that real-world deployment at scale may surface challenges not captured in our controlled evaluation. To address this, we plan to conduct a larger-scale real-world deployment to evaluate \sys{}'s robustness in naturalistic settings. By open-sourcing \sys{}, we will work with the research community to iteratively identify and resolve issues across diverse use cases.

\subsection{Practical Deployment Constraints}

\sys{} can be terminated by the system after periods of inactivity, potentially leading to data loss---a known issue for mobile data collectors~\cite{ferreira_aware_2015}. We implemented a heartbeat mechanism to partially address this and plan to refine \sys{}'s notification system for scenarios where background processes are killed. Additionally, system or app updates can change UI hierarchies, invalidating pre-defined Graph Queries and requiring researchers to update them. We proposed workarounds for similar situations in \cref{table:limitations}. Multi-cultural data collection remains challenging for apps with localization variations (e.g., RTL/LTR layouts) or diverse screen sizes. However, Graph Query provides an extensible foundation to address these issues---for example, using virtual device emulators to generate multiple queries tailored to different locales and configurations.

\subsection{Data Collection Accuracy and Media Type}

Ensuring data collection accuracy without ground truth is challenging, as discussed in \cref{sec:eval-study-two}. Researchers must manually test Graph Queries across multiple screens to verify consistent results. Future work could incorporate end-to-end testing frameworks to automatically validate queries before deployment. Additionally, Android Accessibility Service occasionally captures invisible UI elements, introducing noise that could be mitigated by triangulating with OCR results. Meanwhile, \sys{}'s current focus is on textual data, with potential for richer multimodal data such as screenshots and user interactions. Information rendered in game or 3D engines cannot be collected, as they do not appear in Accessibility Service responses. We plan to integrate additional media types supported by Android Accessibility Service, such as images and screenshots~\cite{android_accessibility_screenshot}.

\subsection{Future Enhancements}

Taking inspiration from PBD studies~\cite{lieberman2000your, cypher1991eager}, future work could ask users to demonstrate multiple target data examples. Developing visualization dashboards would enable real-time monitoring for researchers and participants. Empowering users to delete collected data after dropout would enhance agency. We plan to open source \sys{} to promote ethical use in academic research and welcome community contributions to make it more robust and adaptive to diverse data collection scenarios.
}

\section{Conclusion}

In this paper, we introduce \sys{}, a novel mobile data collection tool that enables researchers to easily define and collect user interface screen data from participants' Android devices. \sys{} utilizes Graph Query, which allows flexible identification, location, and extraction of target UI elements. The tool's key feature is its low-code, programming by demonstration approach that lets researchers specify data to collect through simple interactions with example app screens. A user study with 7 participants demonstrated \sys{}'s ability to effectively collect screen UI data across different devices, while identifying areas for future improvements. Overall, \sys{} represents an important step towards democratizing mobile app data for research by reducing technical barriers and giving participants more agency. Code for \sys{} will be open-sourced to support the academic community in future research data collection.

\begin{acks}
We extend our sincere appreciation to our participants for their contributions to this project. We thank all anonymous reviewers for their feedback. This work is supported in part by the U.S. National Science Foundation under grants CMMI-2326378 and CNS-2426395, a Google Research Scholar Award, and a gift from Adobe Inc. Any opinions, findings, and conclusions or recommendations expressed in this material are those of the authors and do not necessarily reflect the views of the sponsors.

\end{acks}

\balance
\bibliographystyle{ACM-Reference-Format}
\bibliography{references}

\appendix
\section{Appendix}
\subsection{Details of Our Graph Query Translation Prompt}
\label{appendix-prompt}
As described in \cref{sec:query-creation}, we used large language models to translate the candidate graph queries to natural language. Specifically, our prompt uses the following structure:

We defined a Graph Query to locate target data on mobile UI structure. It describes the unique attributes of the data we are targeting. For example, the query 

\begin{lstlisting}
(conj (HAS\_CLASS\_NAME android.widget.FrameLayout) (RIGHT (conj (hasText 6) (HAS\_CLASS\_NAME android.widget.TextView) (HAS\_PACKAGE\_NAME com.ubercab)) ) (HAS\_PACKAGE\_NAME com.ubercab))
\end{lstlisting}

stands for: the information that is located to the right of a text ``6''

Below I have a few queries, can you help me translate them to human-readable format like above?

\begin{enumerate}
    \item \ldots (first query)
    \item \ldots (second query)
    \item \ldots (etc.)
\end{enumerate}

Leave out UI element names (TextView, FrameLayout) and do not make any reference to ``view''. Users only care about the data and information contained in the view instead of the UI elements themselves. ``With numeric index xx'' should be translated into ``the xx in the list'' (first, second, etc.). Be as concise as possible. Make sure you return the translation in the order I presented the queries above, separated by new lines. Return nothing else.








\subsection{Evaluation Study 1 Participant Demographics}
\label{appendix-eval-one-demo}
Here we present the demographics of our evaluation Study 1 participants, as discussed in \cref{sec:eval-one}.
\hyphenpenalty=10000

\begin{table*}[!htbp]  
\small
\begin{tabularx}{\textwidth}{l l l l p{4cm} p{4cm} l}
\toprule
\textbf{ID} & \textbf{Age} & \textbf{Gender} & \textbf{Research Experience} & \textbf{Collected Data Types} & \textbf{Data Collection Tools} & \textbf{Data Collection Projects} \\
\midrule
P1 & 23 & Male   & 3 years & IDE logs, eye tracking, semi-structured interview, survey & IDE plugin, screen recorder, audio recorder, eye tracker & 2 \\
P2 & 30 & Female & 5 years   & app usage, surveys, interviews & Google Analytics, Amazon Web Services, think-aloud, interviews, surveys, Qualtrics, MAXQDA & \ About 6 (5--10) \\
P3 & 26 & Male   & 4 years         & UI interaction, think-aloud, interview & Computer program, recorder & 4 \\
P4 & 32 & Female & 2 years   & Interviews & Audio recording& 2 \\
P5 & 26 & Female & 3.5 years & Gaze, image description & Eye tracker & 2 \\
\bottomrule
\end{tabularx}
\caption{The demographics of our evaluation Study 1 participants, as discussed in \cref{sec:eval-one}.}
\Description{This table summarizes demographic and research background information for five participants in Evaluation Study 1. Each row represents a participant (P1 to P5), and columns include age (ranging from 23 to 32), gender (three female, two male), and years of research experience (from 2 to 5 years). The table also lists the types of data each participant has collected in past projects, such as IDE logs, eye tracking, interviews, surveys, and gaze data. Tools used for data collection include screen recorders, audio recorders, IDE plugins, eye trackers, and online survey platforms. The final column reports the number of data collection projects each participant has worked on, ranging from 2 to about 6.}
\label{tab:eval-one-demo}
\end{table*}
\exhyphenpenalty=10000

\subsection{Evaluation Study 2 Participant Demographics}
\label{appendix-eval-two-demo}
Here we present the demographics of our evaluation Study 1 participants, as discussed in \cref{sec:eval-study-two}.

\begin{table*}[!htbp]
\small
\begin{tabularx}{\textwidth}{XXXlll}
\toprule
\textbf{ID} & \textbf{Age} & \textbf{Gender} & \textbf{Instagram Frequency} & \textbf{Uber Frequency} & \textbf{Chrome Mobile Frequency} \\
\midrule
PB1  & 21  & Male   & Every week             & Once in a few weeks    & Everyday \\
PB2  & 25  & Male   & Everyday               & Once in a few months   & Everyday \\
PB3  & 23  & Male      & Every week             & Every week             & Everyday \\
PB4  & 30  & Male   & Everyday               & Once in a few weeks    & Everyday \\
PB5  & 26 & Female & Everyday               & Once in a few weeks    & Everyday \\
\bottomrule
\end{tabularx}
\caption{Demographics information and app usage frequency for our evaluation Study 2 participants in \cref{sec:eval-study-two}.}
\Description{This table presents demographic and app usage frequency data for five participants in Evaluation Study 2. Each row corresponds to a participant (PB1 to PB5), with columns for age (from 21 to 30), gender (three male, two female), and self-reported frequency of using three mobile apps: Instagram, Uber, and Chrome. Instagram usage ranges from ``Every week'' to “Everyday,'' Uber usage varies from “Every week'' to “Once in a few months,'' and all participants reported using Chrome on mobile “Everyday.'' The table provides context on participants’ familiarity with the apps used during the study.}
\label{tab:eval-two-demo}
\end{table*}

\clearpage

\end{document}